\documentclass[aps,nofootinbib,prd,superscriptaddress,tightenlines,notitlepage,twocolumn,showpacs,floatfix]{revtex4-1}
\usepackage[colorlinks]{hyperref}
\usepackage{tabularx}
\usepackage{bm}
\usepackage{graphicx}
\usepackage{amssymb,bm,tensor}
\usepackage{xcolor}
\usepackage{amsmath}
\usepackage[varg]{txfonts}
\usepackage{enumerate}
\usepackage[normalem]{ulem}
\usepackage{bm}% bold math
\usepackage[OT1]{fontenc}
\usepackage{scalerel,stackengine}
\stackMath
\newcommand\reallywidehat[1]{%
\savestack{\tmpbox}{\stretchto{%
  \scaleto{%
    \scalerel*[\widthof{\ensuremath{#1}}]{\kern-.6pt\bigwedge\kern-.6pt}%
    {\rule[-\textheight/2]{1ex}{\textheight}}%WIDTH-LIMITED BIG WEDGE
  }{\textheight}%
}{0.5ex}}%
\stackon[1pt]{#1}{\tmpbox}%
}

\newcommand{\ZX}[2]{{{#2}}} % Zexing Hu
\begin{document}

\title{Prospects for probing small-scale dark matter models with pulsars around Sagittarius A*}

\date{\today}

\author{Zexin Hu}%\email{huzexin@pku.edu.cn}
\affiliation{Department of Astronomy, School of Physics, Peking University, Beijing 100871, China}
\affiliation{Kavli Institute for Astronomy and Astrophysics, Peking University, Beijing 100871, China}

\author{Lijing Shao}
	\email{lshao@pku.edu.cn}
\affiliation{Kavli Institute for Astronomy and Astrophysics, Peking University, Beijing 100871, China}
\affiliation{National Astronomical Observatories, Chinese Academy of Sciences, Beijing 100012, China}
\affiliation{Max-Planck-Institut f\"ur Radioastronomie, Auf dem H\"ugel 69, D-53121 Bonn, Germany}

\author{Fupeng Zhang}
\affiliation{School of Physics and Materials Science, Guangzhou University, Guangzhou 510006, China}
\affiliation{Key Laboratory for Astronomical Observation and Technology of Guangzhou, 510006 Guangzhou, China}
\affiliation{Astronomy Science and Technology Research Laboratory of Department of Education of Guangdong Province, Guangzhou 510006, China}

\begin{abstract}
  Future observations with next-generation large-area radio telescopes are
  expected to discover radio pulsars (PSRs) closely orbiting around
  Sagittarius~A* (Sgr~A*), the supermassive black hole (SMBH) dwelling at our
  Galactic Center (GC). Such a system can provide a unique laboratory for
  testing General Relativity (GR), as well as the astrophysics around the GC. In
  this paper, we provide a numerical timing model for  PSR-SMBH systems based on
  the post-Newtonian (PN) equation of motion, and use it to explore the
  prospects of measuring the black hole (BH) properties with pulsar timing. We
  further consider the perturbation caused by the dark matter (DM) distribution
  around Sgr~A*, and the possibility of constraining DM models with PSR-SMBH
  systems.  Assuming a 5-year observation of a normal pulsar in an eccentric
  ($e=0.8$) orbit with an orbital period $P_b = 0.5\,$yr, we find that---with
  weekly recorded times of arrival (TOAs) and  a timing precision of 1\,ms---the
  power-law index of DM density distribution near the GC can be constrained to
  about 20\%. Such a measurement is comparable to those measurements at the
  Galactic length scale but can reveal small-scale properties of the DM.
\end{abstract}
%\pacs{}

\maketitle

%% === main body of the paper ===

\allowdisplaybreaks

%---------------------------------------------------------------------
\section{Introduction}
%---------------------------------------------------------------------

Black holes (BHs) are among the most fantastic objects in the Universe.  They
are holding important clues to some open questions in fundamental physics,
concerning the curved spacetime, also possibly the connection of gravitation to
the quantum world~\cite{Hawking:1973uf, Chandrasekhar:1983, Chrusciel:2012jk,
Unruh:2017uaw}.  From astrophysical observations, it is believed that
supermassive BHs (SMBHs) exist at the center of most of massive
galaxies~\cite{McConnell:2012hz}. They provide us with precious opportunities to
probe new physics beyond the current paradigm~\cite{Wex:1998wt, Will:2007pp}.
The SMBH residing in our Galactic Center (GC), Sgr A*, has been confirmed with
observations of S-star orbits as well as the image of its
shadow~\cite{Ghez:2008ms, Genzel:2010zy, EventHorizonTelescope:2022wkp,
EventHorizonTelescope:2022xqj}. As a BH with mass around $4.3\times
10^{6}\,M_{\odot}$ and at a distance to the Solar System of about 8
kpc~\cite{Genzel:2010zy}, Sgr A* has the largest mass to distance ratio among
the known BHs. Thus it is an ideal laboratory for precision BH physics.

In the General Relativity (GR), it is known that for an isolated stationary BH,
the spacetime around it is totally determined by three parameters: the BH mass
($M_{\bullet}$), spin ($S_{\bullet}$) and electric charge. It is the so-called
no-hair theorem~\cite{Israel:1967wq, Carter:1971zc, Robinson:1975bv}.  All
uncharged BHs in GR, which are usually considered in astrophysics, satisfy the
Kerr solution. As a result of the no-hair theorem, all the higher-order multiple
moments of a Kerr BH can be expressed via its mass and
spin~\cite{Hansen:1974zz}. In particular, there is a relation between the BH's
dimensionless spin parameter, $\chi_\bullet \equiv cS_\bullet/GM_\bullet^2$, and
its dimensionless quadrupole moment, $q_\bullet \equiv c^4 Q/ G^2 M_\bullet^3$,
via~\cite{Thorne:1980ru},
%--
\begin{equation}
	q_\bullet=-\chi^2_\bullet\,. \label{eq:nohairtheorem}
\end{equation}
%--
Therefore an independent measurement of a BH's mass, spin and quadrupole can
give a direct test of GR (see e.g. Refs.~\cite{Kramer:2004hd, Liu:2011ae,
Shao:2014wja, Zhang:2017qbb, Bower:2018mta, Weltman:2018zrl, Eatough:2023tst}). 

The mass of Sgr A* can be measured with monitoring gases or stars orbiting it in
close orbits~\cite{Ghez:2008ms,Genzel:2010zy}. Observations of the S2-star's
orbit and its spectroscopy in the last three decades have not only provided a
measurement of the BH mass but also given a clear evidence of the Schwarzschild
precession~\cite{GRAVITY:2020gka}, which sets an upper limit on the extended
mass inside the S2's apocenter of about $3000 \, M_\odot$~\cite{GRAVITY:2021xju,
Heissel:2021pcw}. However, due to the large orbital radius of S2 and the complex
environment near the GC, the spin or quadrupole of Sgr A* may be hard to be
measured with near-future S2 observations~\cite{Merritt:2009ex}. Another effort
that has been done is resolving the shadow of Sgr
A*~\cite{EventHorizonTelescope:2022wkp}, which is a test of the BH metric at the
scale of several Schwarzschild radii. Using the size and shape of the shadow
image, the first set of results from the Event Horizon Telescope Collaboration
gave a consistent constraint on the mass of Sgr A* with that from
S-stars~\cite{EventHorizonTelescope:2022wkp, EventHorizonTelescope:2022exc} and
constrained BH alternatives~\cite{EventHorizonTelescope:2022xqj, Xu:2022frb,
Xu:2023xqh}.

A powerful tool of measuring the properties of the BH spacetime around Sgr A*
is observing a pulsar orbiting it in a close orbit~\cite{Zhang:2017qbb,
Liu:2011ae, Psaltis:2015uza, Dong:2022zvh}. Taking the advantage of the high
accuracy of the pulsar timing technology, it is expected to detect the spin and
quadrupole moment of Sgr A* with proper pulsars~\cite{Liu:2011ae,
Psaltis:2015uza, Bower:2018mta}. Searches of pulsars in the GC have been carried
out several times over the last few decades but no pulsar \ZX{}{ within the inner parsec} has been found yet~\cite{Kramer:2000tc, Eatough:2013nva,
Wharton:2011dv, Siemion:2013, Liu:2021ziv, EHT:2023hcj}. Although observational evidence and
theoretical model both indicate that there could be a number of neutron stars in
the GC~\cite{Wharton:2011dv}, the large dispersion measures and highly turbulent
interstellar medium in the GC region make the detection nearly impossible at the
typical low-frequencies~\cite{Cordes:2002wz}. The observational sensitivity at
higher frequencies are limited by the steep spectrum of pulsar emission.
Nevertheless, future observations with the next-generation telescopes, such as
the Square Kilometre Array (SKA) and the next-generation Very Large Array
(ngVLA), are still hopeful of finding those pulsars~\cite{Liu:2011ae,
Bower:2018mta}. Such a discovery could open a new avenue of testing
gravity~\cite{Kramer:2004hd, Shao:2014wja}. 

Because of the potential for breakthrough, it is crucial to develop a pulsar
timing model for  PSR-SMBH systems. Several pioneering works~\cite{Liu:2011ae,
Psaltis:2015uza} have been done based on an analytic solution that consistently
includes the periodic spin effects derived by \citet{Wex:1995}, which can be
regarded as an elegant extension of the widely used Damour-Deruelle (DD) timing
model~\cite{Damour:1986} for this particular situation. However, it is in
general hard to analytically include both the spin and quadrupole effects
simultaneously, and no elegant solution has be found yet. Nevertheless, the
influences caused by the quadrupole effect can be included as perturbations to
the pulsar coordinate position as well as for an additional secular
precession~\cite{Liu:2011ae}. In this paper, we are going to develop a
preliminary pulsar timing model based on the direct numerical integration of the
post-Newtonian (PN) equation of motion that includes the spin-orbital coupling
and quadrupole interaction. Comparing to applying a fully GR calculation based
on the Kerr metric~\cite{Zhang:2017qbb}, the PN formalism allows us to treat the
BH's spin and quadrupole moment as independent variables, which is favored when
testing alternative gravity theories. The numerical method is also more flexible
when considering new contributions from other perturbations. Different from the
analytic approach, which describes the system with parameters represented for
separate effects, such as the periastron advance parameter, $k$, and the
deformation parameters of the orbit, $\delta_r$ and $\delta_\theta$, in the DD
model~\cite{Damour:1986}, the numerical method uses the true physical parameters
that directly related to the system, which is more convenient when further
extending the model.

As we have mentioned before, the complex environment near the GC may spoil the
PSR-SMBH system. The angular momentum and quadrupole moment contributed by the
stellar cluster can obscure the signal caused by the BH's spin and quadrupole,
which requires the orbit of the pulsar to be inside $0.1$--$1$ mpc in order to
meaningfully measure the BH's properties~\cite{Psaltis:2015uza, Merritt:2009ex}.
Moreover, a high fraction of objects with mass about $10\,M_\odot$ may make the
tests of gravity problematic at all radii~\cite{Merritt:2009ex}. A conservative
method is to only use the timing data that the pulsar is around the periastron,
where the SMBH dominates the orbital dynamics of the
pulsar~\cite{Psaltis:2015uza}.  Another possible approach is to include the main
contribution of the external perturbations as part of the timing model and do
parameter estimation simultaneously. Similar idea has been used to constrain the
existence of intermediate-mass BH in the GC with the S2
orbit~\cite{GRAVITY:2023met}.

In this paper we consider another kind of important perturbation in PSR-SMBH
systems, that is the dark matter (DM) distribution around the SMBH. In the
standard Lambda Cold Dark Matter (${\rm \Lambda CDM}$) model, the galaxies are
formed inside DM halos, which consist of DM particles/fields that account for
about 27\% of the mass-energy at present universe but with a yet unknown
nature~\cite{Planck:2015fie}.  While the DM distributions at large scales can be
measured with galactic rotation curves or gravitational lensing, the small scale
structures have seldom been constrained by observations~\cite{Hui:2016ltb}.
Cold-DM-only simulation showed that the DM density has a cusp $\propto 1/r$  in
the halo center, which is a known feature of the so-called Navarro-Frenk-White
(NFW) profile~\cite{Navarro:1996gj}.  However, baryonic processes like the
adiabatic growth of the central BH or supernova feedback can modify the profile,
while some other possible DM models, such as the ultralight DM model, also
predict different central density distributions~\cite{Hui:2016ltb}. Observations
of the galactic rotation curve have shown both evidences for cusp and core-like
distributions~\cite{Moore:1999gc}. A cuspy density profile around the GC may
contribute enough DM mass inside the pulsar obit that can be detected by pulsar
timing observation, while satisfies the constraints from the S2
observation~\cite{Heissel:2021pcw, Lacroix:2018zmg, shen2023exploring,
Zakharov:2007fj}. \ZX{}{Recent work using the high-precision timing results of millisecond binary pulsars to directly measure the Galactic acceleration and derive fundamental Galactic parameters also shows the potential of detecting the DM distribution with pulsar timing technique~\cite{Chakrabarti:2020abx}.} In this paper, we explore the prospects of constraining DM
models via timing a pulsar around Sgr A* by extending our timing model to
include the DM contribution. We consider the DM perturbation as a spherical mass
distribution at the leading order and ignore the triaxial deformation of the DM
that may exist in the CDM model~\cite{Dubinski:1991bm}.

The remaining part of this paper is organized as follows. In
Sec.~\ref{sec:orbital dynamics}, we describe the orbital dynamics of a PSR-SMBH
system based on the PN equation of motion. Section~\ref{sec:pulsar timing}
presents the basic concept of pulsar timing as well as the various effects we
take into consideration in our timing model. We give an inverse timing formula
of this numerical timing model in Sec.~\ref{sec:inv timing} and use it in
Sec.~\ref{sec:para esti} to get the expected measurement precision of the BH
properties. In Sec.~\ref{sec:DM pert}, we extend our timing model to include the
effects of DM distribution around the GC and show the parameter estimation
results obtained from the extended timing model. Finally, we conclude in
Sec.~\ref{sec:conclusion}.

%---------------------------------------------------------------------
\section{Orbital dynamics}\label{sec:orbital dynamics}
%---------------------------------------------------------------------

Differently from the Newtonian case, the two-body problem in GR has no general
analytic solution. One can obtain the approximate equations of motion for well
separated systems through the so-called PN expansion~\cite{Damour:1985}. We
invoke PN equation of motion for the two-body orbital
dynamics~\cite{Barker:1975ae, Kidder:1995zr, Blanchet:2013haa}, 
%--
\begin{equation}\label{eq:PN:acc}
  \ddot{\bm{r}} \equiv \frac{ {\rm d}^2 \bm{r}}{{\rm d}t^2} =
  \ddot{\bm{r}}_{\rm N} + \ddot{\bm{r}}_{\rm 1PN}  +
  \ddot{\bm{r}}_{\rm SO} + \ddot{\bm{r}}_{\rm Q} + \ddot{\bm{r}}_{\rm 2PN}+ \ddot{\bm{r}}_{\rm 2.5PN} +
  \cdots \,,
\end{equation}
%--
where $\bm{r} \equiv \bm{r}_{\rm PSR} - \bm{r}_\bullet$ is the relative
coordinate position vector in the harmonic gauge, and $t$ is the coordinate
time. By writing down the above equation, we have restricted ourselves to the
case that only the BH is spinning, as the pulsar spin has, in general,
negligible effects on the orbital motion. The Newtonian acceleration in
Eq.~(\ref{eq:PN:acc}) reads $\ddot{\bm{r}}_{\rm N} = - G M \hat{\bm{n}} / r^2$,
with $M \equiv M_\bullet + m_{\rm PSR}$, $r \equiv \left|\bm{r}\right|$,  and
$\hat{\bm{n}} \equiv \bm{r} / r$.

Besides the Newtonian term, other terms in Eq.~(\ref{eq:PN:acc}) represent
contributions from higher-order PN terms, spin-orbit coupling, quadrupolar
effects to the orbit, and so on.  For our purpose, we consider the following
terms, 
%---
\begin{widetext}
\begin{subequations}\label{eq:PN:acc}
\begin{eqnarray}
  \ddot{\bm{r}}_{\rm 1PN} &=& - \frac{GM}{c^2r^2} \left\{ \left[ \left(
  -4-2\eta \right) \frac{GM}{r} + \left( 1+3\eta \right) v^2 - \frac{3}{2}\eta
  \dot r^2 \right] \hat{\bm{n}} - \left(  4-2\eta \right) \dot r \bm{v}
  \right\} \,, \\
  %---
  \ddot{\bm{r}}_{\rm SO} &=& \chi_\bullet \frac{G^2M^2}{4c^3r^3} \left(1 +
  \sqrt{1-4\eta}\right) \left\{ \left[ 12 \hat{\bm{s}} \cdot \left(
  \hat{\bm{n}} \times \bm{v} \right) \right] \hat{\bm{n}} + \left[ \left( 9 +
    3\sqrt{1-4\eta} \right) \dot r \right] \left( \hat{\bm{n}} \times
    \hat{\bm{s}} \right) - \left( 7+\sqrt{1-4\eta} \right) \left( \bm{v} \times
    \hat{\bm{s}} \right) \right\} \,, \\
  %---
  \ddot{\bm{r}}_{\rm Q} &=& - q_\bullet \frac{3G^3 M_\bullet^2 M}{2c^4 r^4} \left\{
    \left[ 5\left( \hat{\bm{n}} \cdot \hat{\bm{s}} \right)^2 - 1 \right]
    \hat{\bm{n}} - 2\left( \hat{\bm{n}} \cdot \hat{\bm{s}} \right) \hat{\bm{s}}
  \right\} \,,
\end{eqnarray}
\end{subequations}
\end{widetext}
%---
where $\dot r \equiv {\rm d}r/{\rm d}t$, $\bm{v} \equiv {\rm d}\bm{r} / {\rm
d}t$, $v \equiv \left| \bm{v} \right|$, $\eta \equiv m_{\rm PSR} M_\bullet /
M^2$, and $\hat{\bm{s}}$ is the unit vector pointing along the BH spin.  In
particular, in our simulation the acceleration at 2PN order, $\ddot{\bm{r}}_{\rm
2PN}$, which is smaller by ${\cal O} \left(v^2/c^2\right)$ than
$\ddot{\bm{r}}_{\rm 1PN}$, and the acceleration at 2.5PN order, which leads to
the emission of gravitational waves, are not included~\cite{Blanchet:2013haa}.
The acceleration caused by the BH's quadrupole is numerically at the 2PN order.
For consistency we should include at least the $\ddot{\bm{r}}_{\rm 2PN}$ term,
but our treatment could be enough for the purpose of {\it forecasting} the
precision of measuring the BH properties.  In {\it measuring} the parameters
from real data, more terms are in demand.  We leave this for future studies.

In addition, because the mass ratio $m_{\rm PSR} / M_\bullet < 10^{-6}$, we will
ignore the mass of the pulsar for the moment.  By doing this, we do not consider
the back-reaction of the orbit to the spin of the BH, which is proportional to
$\eta$~\cite{Barker:1975ae}; thus the spin of the BH stays constant in our
simulation and the pulsar is moving like a test particle in the BH's spacetime.
Also, we will have a constant position for the BH that, $\dot{\bm{r}}_\bullet =
0$ and $\bm{v} = \bm{v}_{\rm PSR}$.  With $\eta \simeq 0$, Eq.~(\ref{eq:PN:acc})
simplifies to,
%---
\begin{subequations}\label{eq:PN:acc:simple}
\begin{eqnarray}
  \hspace{-0.5cm}
  \ddot{\bm{r}}_{\rm 1PN} &\simeq& - \frac{GM_\bullet}{c^2r^2} \left[ \left( -
  \frac{4GM}{r} +  v^2  \right) \hat{\bm{n}} - 4 \dot r \bm{v} \right] \,,\\
  %---
  \hspace{-0.5cm}
  \ddot{\bm{r}}_{\rm SO} &\simeq& \chi_\bullet \frac{6G^2M_\bullet^2}{c^3r^3}  \left[
  \hat{\bm{s}} \cdot \left( \hat{\bm{n}} \times \bm{v} \right) \hat{\bm{n}} + \dot r  \left( \hat{\bm{n}} \times \hat{\bm{s}} \right) -\frac{2}{3} \left(
  \bm{v} \times \hat{\bm{s}} \right) \right] \,, \nonumber\\
  && \\
  %---
  \hspace{-0.5cm}
  \ddot{\bm{r}}_{\rm Q} &\simeq& - q_\bullet \frac{3G^3 M_\bullet^3 }{2c^4 r^4} \left\{
    \left[ 5\left( \hat{\bm{n}} \cdot \hat{\bm{s}} \right)^2 - 1 \right]
    \hat{\bm{n}} - 2\left( \hat{\bm{n}} \cdot \hat{\bm{s}} \right) \hat{\bm{s}}
  \right\} \,.
\end{eqnarray}
\end{subequations}
%---

Our approximations are all well justified if our purpose is only to assert the
precision of measuring the mass, spin, and quadrupole of the Sgr A* via timing a
pulsar around it. When one wants to conduct a practical timing model to fit the
real time-of-arrival (TOA) data from such a system, depending on the orbital
characteristics, one needs to reconsider these approximations and to take into
account all terms that would lead to timing residuals larger than the noises in
observation. In general, by including higher-order terms one may have better
power to break degeneracy among parameters due to the varieties that would be
introduced by these terms. In this regard, our treatment is on the conservative
side. The same arguments apply to the treatment of the Einstein delay and the
Shapiro delay in Sec.~\ref{sec:pulsar timing}. We wish to track down the effects
from higher-order terms in future investigation.

%---
\begin{figure}
  \centering
  \includegraphics[width=9cm, trim=0 50 0 15]{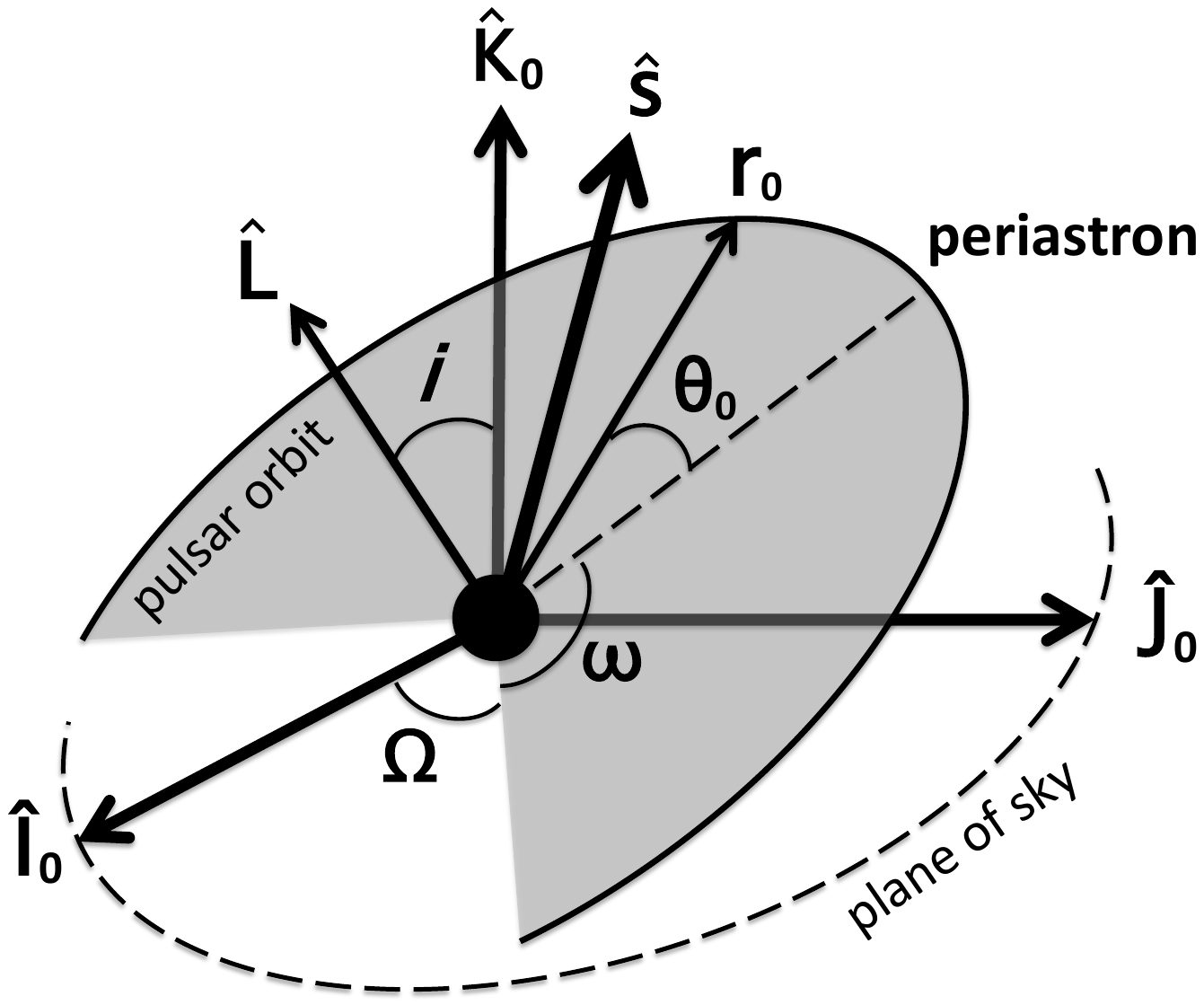}
  \caption{\label{fig:geometry} Coordinate system and notations that are used
  for a pulsar orbit around Sgr A*. The frame $\left( \hat{\bm{I}}_0,
  \hat{\bm{J}}_0, \hat{\bm{K}}_0  \right)$ has $\hat{\bm{K}}_0$ pointing from
  the Earth towards Sgr A*, and $\left( \hat{\bm{I}}_0, \hat{\bm{J}}_0\right)$
  constitutes the sky plane with $\hat{\bm{I}}_0$ pointing towards east and
  $\hat{\bm{J}}_0$ towards north. The orientation of the pulsar orbit is
  determined by the inclination, $i$, the longitude of periastron, $\omega$, and
  the longitude of ascending node, $\Omega$. The Sgr A* sits at the origin, and
  its spin points towards $\hat{\bm{s}} \equiv \left( \sin\lambda_\bullet
  \cos\eta_\bullet, \, \sin\lambda_\bullet\sin\eta_\bullet, \,
  \cos\lambda_\bullet \right)$ in the coordinate system; in the figure
  $\lambda_\bullet$ and $\eta_\bullet$ are not shown for clarity. The position
  of the pulsar at the reference time $t=0$ is $\bm{r}_0$, and it is described
  by the true anomaly $\theta_0$. }
\end{figure}
%---

The orbit of a pulsar around Sgr A* is described by the orbital period, $P_b$,
the eccentricity, $e$, and various angles that determine the orientaion; see
Fig.~\ref{fig:geometry} for definition of these angles. A Keplerian description
of the orbit is understood to be an approximation to the instantaneous motion of
the pulsar. In our simulation, the orbital elements are updated according to the
instantaneous position, $\bm{r}$, and instantaneous velocity, $\bm{v}$. For
example, the eccentricity, rigorously speaking, should be a function of time,
$e(t)$, due to various non-Newtonian acceleration terms in
Eq.~(\ref{eq:PN:acc}). If not mentioned explicitly, the values of orbital
elements, for example in Eq.~(\ref{eq:example:orbit}), refer to the reference
time $t=0$ in the simulation.

As a fiducial case, we study a PSR-Sgr A* system with the following
parameters,
%---
\begin{subequations}\label{eq:example:orbit}
\begin{eqnarray}
  M_\bullet &=& 4.3 \times 10^6 \, {\rm M}_\odot  \label{eq:example:orbit:M} 
  \,,\\
  \chi_\bullet &=& 0.6 \,,  \quad \lambda_\bullet = \frac{1}{3} \pi \,, \quad
  \eta_\bullet = \frac{5}{9} \pi \,, \quad q_\bullet = -0.36
  \label{eq:example:orbit:spin} \,,\\
  P_b &=& 0.5 \, {\rm yr} \,, \quad e = 0.8 \label{eq:example:orbit:orbit}
  \,,\\
  i &=& \frac{1}{5} \pi \,, \quad \omega = \frac{5}{7} \pi \,, \quad \theta_0 =
  \frac{1}{3} \pi \,. \label{eq:example:orbit:angles}
\end{eqnarray}
\end{subequations}
%---
The angles are quasi-randomly chosen, to avoid special orientations that might
render the parameter-estimation problem degenerate.\footnote{The value of
$\lambda$ was inspired by \citet{Psaltis:2014dea}.} We integrate the orbit for
$t_{\rm total} = 5\,$yr, that contains $t_{\rm total}/P_b \simeq 10$ orbits.
TOAs are extracted weekly and uniformly in time. For 5 years, we simulate 260
TOAs in total. In the simulation, the longitude of the ascending node is fixed
to $\Omega = 0$.  This choice is generic, because in this case the rotation of
the pulsar orbit around $\hat{\bm{K}}_0$ does not affect TOAs if we ignore the
proper motion of the Sgr~A*~\cite{Taylor:1994zz}.

%---
\begin{figure}
  \centering
  \includegraphics[width=8.5cm, trim=40 20 0 30]{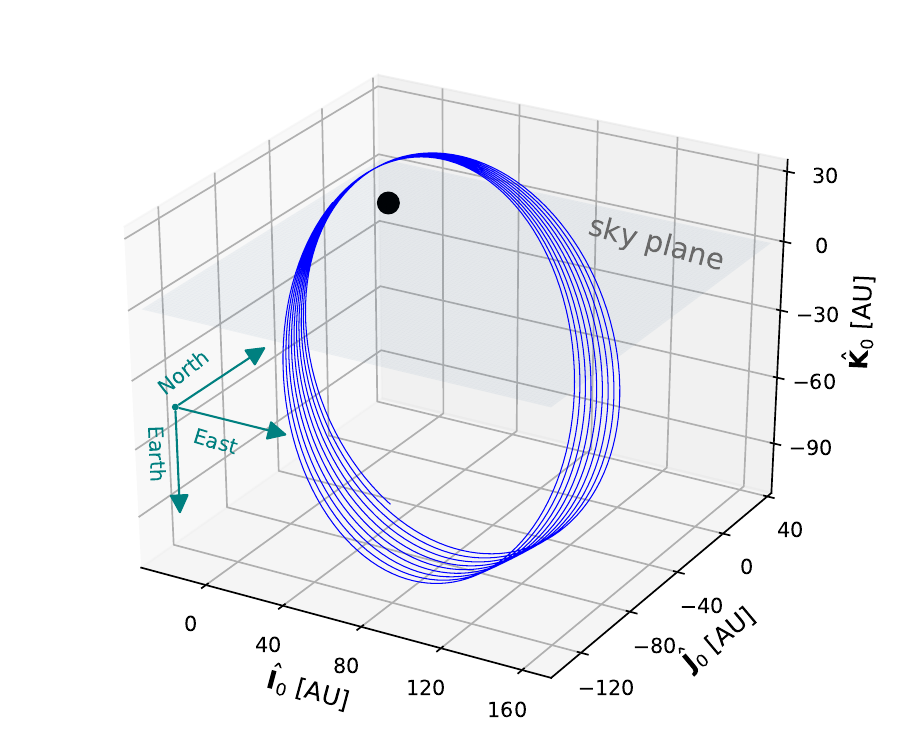}
  \caption{Illustration of the pulsar orbits around Sgr~A* with $P_b=0.5$\,yr,
  $e=0.8$, and $\chi_\bullet=0.6$; see text and Eq.~(\ref{eq:example:orbit}) 
  for more details.  \label{fig:orbit:3d}}
\end{figure}
%---

The orbits with Eq.~(\ref{eq:example:orbit}) are illustrated in
Fig.~\ref{fig:orbit:3d}. It is seen in the figure the precession of the orbits
due to the PN acceleration $\ddot {\bm{r}}_{\rm 1PN}$ and the frame-dragging
effects from the BH spin~\cite{Lense:1918zz, Schiff:1960gh}.

%---------------------------------------------------------------------
\section{Pulsar timing}\label{sec:pulsar timing}
%---------------------------------------------------------------------

In pulsar timing, the TOAs of pulsar pulses at the telescopes are connected to
the proper rotation numbers, $N$, of the pulsar~\cite{Blandford:1976,
Damour:1986, Taylor:1994zz, Shao:2022izp, Hu:2023vsq}. The timing model
incorporates the various effects in the orbital motion and radiation
propagation, and thus from the observed TOAs one can extract the underlying
physical parameters. For the pulsar's proper rotation, we assume,
%---
\begin{equation} \label{eq:psr:spindown}
  N(T) = N_0 + \nu T + \frac{1}{2} \dot \nu T^2 \,,
\end{equation}
%---
where $\nu \equiv 1/P$ is the pulsar's spin frequency, $\dot\nu$ is the time
derivative of $\nu$, and $T$ is the proper time of the
pulsar~\cite{Damour:1986}.  Higher-order time derivatives are easy to include
when needed.

At the lowest order, the proper time of the pulsar is connected to the
coordinate time $t$ via,
%---
\begin{equation}\label{eq:diff:Einstein}
  \frac{{\rm d} T}{ {\rm d}t} = 1 - \frac{G M_\bullet}{c^2r} - \frac{v^2}{2c^2}
  + \cdots \,.
\end{equation}
%--
Integrating the above equation gives an Einstein delay that accounts for the
gravitational redshift and special-relativistic time-dilation
effects~\cite{Blandford:1976, Damour:1986},
%---
\begin{equation}\label{eq:Einstein}
  \Delta_{\rm E} \equiv t - T \,.
\end{equation}
%---
To drop a term that is linear in time, one can redefine the pulsar
spin~\cite{Damour:1986} and Eq.~(\ref{eq:diff:Einstein}) takes a form as
%---
\begin{equation}\label{eq: Einstein 2}
	\frac{{\rm d}T}{{\rm d}t} = \frac{1-\frac{GM_\bullet}{c^2 r} -
	\frac{v^2}{2c^2}}{\left\langle 1-\frac{GM_\bullet}{c^2
	r}-\frac{v^2}{2c^2}\right\rangle}\,.
\end{equation}
%---
Roughly speaking, the $\left\langle\, \cdot \right\rangle$ term that appears in
the denominator means to average over the pulsar orbital period, so that
$\left\langle {\rm d}T/{\rm d}t\right\rangle=1$ and there is no linear-in-time
dependence term in the Einstein delay. Due to the spin and quadrupole effects,
the denominator is no longer a constant. Here we use the 1PN approximation of
the denominator, which accounts for almost all the linear dependence.

For a Keplerian orbit, $\Delta_{\rm E}$ is found to be~\cite{Blandford:1976,
Damour:1986},
%---
\begin{equation}
  \Delta_{\rm E} = \frac{2e}{c^2} \left( \frac{G^2M_\bullet^2 P_b}{2\pi} 
  \right)^{1/3} \sin u \,,
\end{equation}
%---
where $u$ is the eccentric anomaly. In our simulation, in order to account for
all the factors coming from the variation of the orbit, we integrate
Eq.~(\ref{eq: Einstein 2}) to obtain the Einstein delay.

The orbital motion of the pulsar gives the geometric delay, called the R\"omer
delay, which is simply,
%---
\begin{equation}\label{eq:Roemer}
  \Delta_{\rm R} \equiv \frac{1}{c} \hat{\bm{K}}_0 \cdot \bm{r} \,.
\end{equation}
%---

We also include the lowest-order propagation time delay caused by the curvature
of the Sgr A*, called the 1PN Shapiro delay~\cite{Shapiro:1964uw,
Blandford:1976},
%---
\begin{equation}\label{eq:Shapiro}
  \Delta_{\rm S} = -\frac{2GM_\bullet}{c^3} \log \left( r - \bm{r} \cdot
  \hat{\bm{K}}_0   \right) \,.
\end{equation}
%---

Finally, we assume that TOAs are collected at an infinite distance to the Sgr
A*. In doing so, we are ignoring various terms related to the proper motion of
the Sgr~A* and the motion of the Earth around the Sun, {\it etc.}. After
dropping the constant (infinite) term, one finally has~\cite{Damour:1986},
%---
\begin{equation}\label{eq:toa}
  t^{\rm TOA} = T + \Delta_{\rm R} + \Delta_{\rm E} + \Delta_{\rm S} \,.
\end{equation}
%---

Notice that we are only considering the lowest-order terms in the Einstein delay
and the Shapiro delay. In principle, higher-order terms can also be added; for
examples, see \citet{Kopeikin:1997} and  \citet{Wex:1998wt} for the higher-order
terms in the propagation delay of pulses. We suspect that the higher-order terms
will further break the degeneracies in the parameter-estimation problem which,
as we mentioned, renders our treatment conservative. We \ZX{}{ defer} the
investigation for a future study.

%---
\begin{figure}
  \centering
  \includegraphics[width=8.5cm, trim=0 20 0 0]{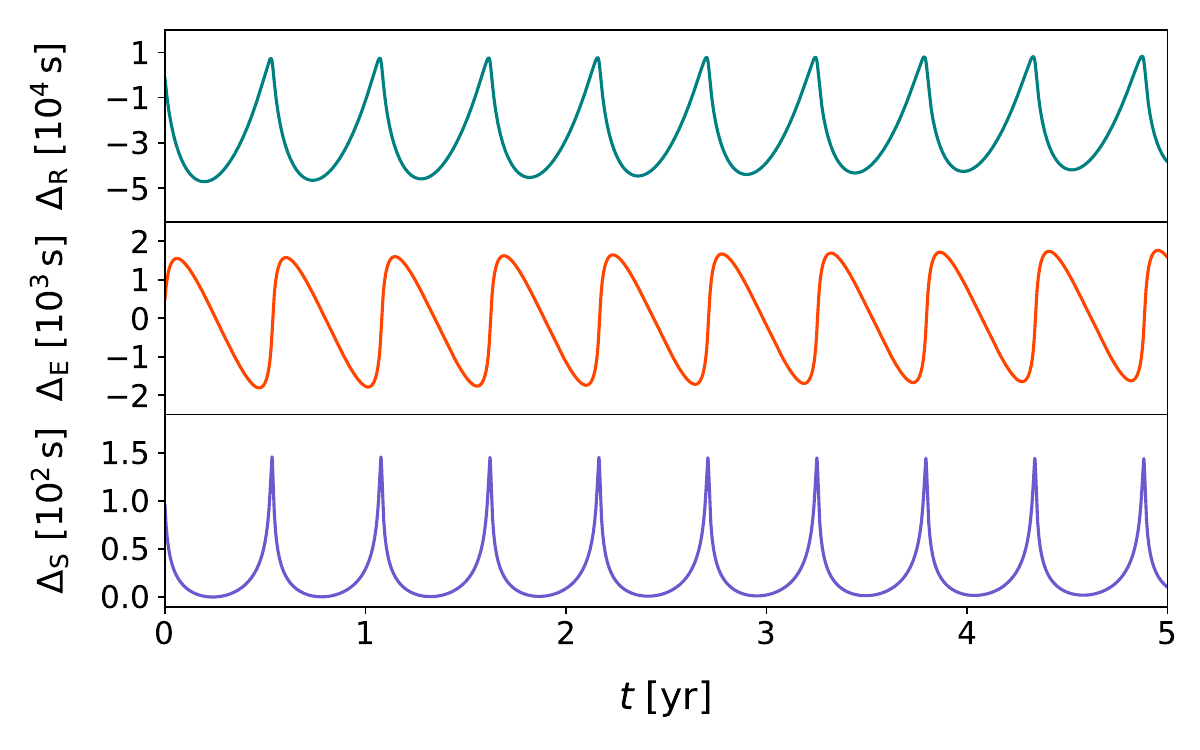}
  \caption{Three time delays in pulsar timing, defined in
    Eq.~(\ref{eq:Roemer}), Eq.~(\ref{eq:Einstein}), and Eq.~(\ref{eq:Shapiro}),
    for the fiducial orbit (\ref{eq:example:orbit}). The linear trend  in time
    in $\Delta_{\rm E}$ is removed; it can be absorbed into a redefinition of
  pulsar's spin.}
  \label{fig:time:delays}
\end{figure}
%---

We give an illustration in Fig.~\ref{fig:time:delays} for these three time
delays for the fiducial orbit defined in Eq.~(\ref{eq:example:orbit}). These
delays are extremely large compared with those of the binary pulsar systems we
are currently regularly timing~\cite{Manchester:2004bp, Liu:2011ae}. Notice that
for our fiducial orbit, the inclination is only $\pi/5 = 36^\circ$. Even with
such a small inclination, the Shapiro delay is already numerous; this is also
true even for face-on orbits~\cite{Liu:2011ae}.

%---------------------------------------------------------
\section{The inverse timing formula}\label{sec:inv timing}
%---------------------------------------------------------

In the standard procedure of parameter estimation in pulsar timing, for
calculating the residuals  one needs an inverse timing formula that calculates
$N$ from given TOAs and system's parameters $\bm{\Theta}$ instead of the pulsar
timing model described in Sec.~\ref{sec:pulsar timing} that calculates $t^{\rm
TOA}(N;\bm{\Theta})$~\cite{Damour:1986}. In principle, giving the system's
parameters $\bm{\Theta}$ and the observed TOA, $t^{\rm TOA}$, we can first
integrate the pulsar's orbital motion and get a series of time delays that are
related to the coordinate time: $\Delta_{\rm R}(t_i)\,,\Delta_{\rm
S}(t_i)\,,\Delta_{\rm E}(t_i)$. Then we can do interpolation for these series
and solve the implicit equation, $t^{\rm TOA}=t+\Delta_{\rm R}(t)+\Delta_{\rm
S}(t)$, to get the coordinate pulse emission time, $t$. Finally we use
$T=t-\Delta_{\rm E}(t)$ and Eq.~(\ref{eq:psr:spindown}) to get the related $N$.
But a problem rised in doing this procedure numerically. To have an acceptable
precision in solving the implicit equation, the interpolation step needs a dense
coverage, which makes the orbital integration slow. So here we propose a
fast method to get the inverse timing formula.

Briefly speaking, we want to change the variable $t$ in the differential
equations~(\ref{eq:PN:acc}) and~(\ref{eq:diff:Einstein}) to $t^{\rm TOA}$ that
is directly related to the observation.

From Eq.~(\ref{eq:Einstein}) and Eq.~(\ref{eq:toa}), we have
%--
\begin{equation}
	t^{\rm TOA}=t+\Delta_{\rm R}(t)+\Delta_{\rm S}(t)\,.
\end{equation}
%--
Taking a derivative on both sides and insert the explicit forms shown in
Eq.~(\ref{eq:Roemer}) and Eq.~(\ref{eq:Shapiro}), we get
%--
\begin{equation}\label{eq:dtoadt}
	\frac{{\rm d}t^{\rm TOA}}{{\rm d}t} = 1+\frac{1}{c}\hat{\bm{K}}_0\cdot
	\bm{v}-\frac{2GM_{\bullet}}{c^3}\frac{\hat{\bm{n}}\cdot\bm{v} -
	\bm{v}\cdot\hat{\bm{K}}_0}{r-\bm{r}\cdot\hat{\bm{K}}_0}\,.
\end{equation}
%--
Combine this with Eq.~(\ref{eq:PN:acc}) and Eq.~(\ref{eq:diff:Einstein}), we can
get the complete differential equations used in the inverse timing formula,
which are
%--
\begin{eqnarray}
	\frac{{\rm d}\bm{r}}{{\rm d}t^{\rm TOA}}&=&\bm{v}\frac{{\rm d}t}{{\rm d}t^{\rm TOA}}\,,\\
	\frac{{\rm d}\bm{v}}{{\rm d}t^{\rm TOA}}&=&\ddot{\bm{r}}\frac{{\rm d}t}{{\rm d}t^{\rm TOA}}\,,\\
	\frac{{\rm d}t}{{\rm d}t^{\rm TOA}}&=&\frac{{\rm d}t}{{\rm d}t^{\rm TOA}}\,,\\
	\frac{{\rm d}\Delta_{\rm E}}{{\rm d}t^{\rm TOA}}&=&\left(1-\frac{{\rm d}T}{{\rm d}t}\right)\frac{{\rm d}t}{{\rm d}t^{\rm TOA}}\,,
\end{eqnarray}
%--
where ${\rm d}t/{\rm d}t^{\rm TOA}$ is the inverse of Eq.~(\ref{eq:dtoadt}).
Note that we also need to transform the initial conditions. From the conditions
that at $t=0$, the system has parameters $\bm{\Theta}$ and $\Delta_{\rm E}=0$,
we can simply calculate $t^{\rm TOA}_0=\Delta_{\rm R}\big|_{t=0} +\Delta_{\rm
S}\big|_{t=0}$. The initial conditions now are at $t^{\rm TOA}=t^{\rm TOA}_0$,
and the system has parameters $\bm{\Theta}$ and $\Delta_{\rm E}=0$.

Integrating the above equations from $t^{\rm TOA}_0$ to $t^{\rm TOA}$, we can
get the related pulsar proper time $T$ by
%--
\begin{equation}
	T=t^{\rm TOA}-\Delta_{\rm R}\big(t^{\rm TOA}\big)-\Delta_{\rm E}\big(t^{\rm
	TOA}\big)-\Delta_{\rm S}\big(t^{\rm TOA}\big)\,,
\end{equation}
%--
and solve $N$ from Eq.~(\ref{eq:psr:spindown}) easily. We will use this method
as the inverse timing formula in the following numerical calculation.

%---------------------------------------------------------------------
%\section{Markov-chain Monte Carlo parameter estimation}
\section{parameter estimation}\label{sec:para esti}
%---------------------------------------------------------------------

After all parameters, which we denote as $\bm{\Theta}$, are given, one obtains
the pulsar rotation number $N$ as a function of $t^{\rm TOA}$ without any
ambigiuity. The parameters include,
%---
\begin{equation}
  \bm{\Theta} = \bm{\Theta}_\bullet \cup \bm{\Theta}_{\rm orbit} \cup
  \bm{\Theta}_{\rm PSR} \cup \bm{\Theta}_{\rm pert} \,,
\end{equation}
%---
where $\bm{\Theta}_\bullet$ includes parameters of the BH in
Eqs.~(\ref{eq:example:orbit:M}--\ref{eq:example:orbit:spin}), $\bm{\Theta}_{\rm
orbit}$ includes parameters of the orbit in
Eqs.~(\ref{eq:example:orbit:orbit}--\ref{eq:example:orbit:angles}),
$\bm{\Theta}_{\rm PSR}$ includes parameters of the pulsar spin in
Eq.~(\ref{eq:psr:spindown}), and $\bm{\Theta}_{\rm pert}$ includes parameters
from the perturbation.  Therefore, we denote the pulsar rotation number as
$N\left(\bm{\Theta}; t^{\rm TOA}\right)$.  $\bm{\Theta}_{\rm pert}$ is to be
introduced in Sec.~\ref{sec:DM pert}, and we omit it in this section.

Here in $\bm{\Theta}_\bullet$, we treat $\chi_\bullet$ and $q_\bullet$ as
independent variables. In GR, the no-hair theorem poses $q_\bullet = -
\chi_\bullet^2$~\cite{Israel:1967wq, Chrusciel:2012jk}. Our treatment can be
viewed as an expansion of the multipoles of the spacetime~\cite{Thorne:1980ru}
with the three lowest-order moments, namely the monopole $M_\bullet$, the dipole
$S = \chi_\bullet GM^2/c$, and the quadrupole $ Q = q_\bullet G^2M^3 /c^4$. The
post-Newtonian expansion in Eq.~(\ref{eq:PN:acc}) provides the possibility to
treat $\chi_\bullet$ and $q_\bullet$ independently, thus providing the
possibility to assess the precision in testing the no-hair theorem.

Assuming a Gaussian timing noise realization in observation, the probability
that the true values of parameters being $\bm{\Theta}$ is,
%---
\begin{equation}\label{eq:prob}
  P\left(\bm{\Theta} \left| t^{\rm TOA} \right.\right) \propto \exp\left(
  -\frac{P^2}{2} \sum_{i=1}^{N_{\rm TOA}} \frac{
    \left[N^{(i)}\left(\bm{\Theta}\right) -
    N^{(i)}\left(\bar{\bm{\Theta}}\right) \right]^2}{\sigma_{\rm TOA}^2}
    \right) \,,
\end{equation}
%---
where $N^{(i)}\left(\bm{\Theta}\right) \equiv N\left(\bm{\Theta}; t^{\rm
TOA}_i\right)$, and the summation is over the number of TOAs. In real data, the
signal is contaminated with noises, and the simulation can take them into
account by adding, say, Gaussian noises, as was done in
Refs.~\cite{Shao:2016ubu, Shao:2017lmp}. Nevertheless, it is statistically
equivalent to use the noiseless templates, $N\left(\bar{\bm{\Theta}}; t^{\rm
TOA}\right)$ in Eq.~(\ref{eq:prob}). The addition of random noises only shifts
$\ln P\left(\bm{\Theta} \left| t^{\rm TOA} \right.\right)$ by a constant value
statistically for the expecting value of an ensemble of noise realizations.  The
use of noiseless templates avoids the randomness in the realization of noises,
and it was also adopted in other scientific studies; see e.g.
\citet{Gaebel:2017zys} for simulations of parameter estimation with
gravitational-wave waveform templates.

As a standard procedure for parameter estimation~\cite{Damour:1986,
Edwards:2006zg}, we estimate the measurement uncertainties of these parameters
via the covariance matrix
%-----
\begin{equation}
	C_{\alpha\beta} = \left(\frac{\partial^2\mathcal{L}}{\partial\Theta^\alpha
	\partial\Theta^\beta}\right)^{-1}\,,
\end{equation} 
%-----
where $\mathcal{L}=-\ln P\big(\Theta|t^{\rm TOA}\big)$ is the log-likelihood
function. We assume the timing precision $\sigma_{\rm TOA}=1\,{\rm ms}$ in our
simulation, which is a relatively conservative estimate for future
observations~\cite{Liu:2011ae}.

%----
\begin{figure}
	\centering
	\includegraphics[width=9.5cm, trim=0 0 0 20]{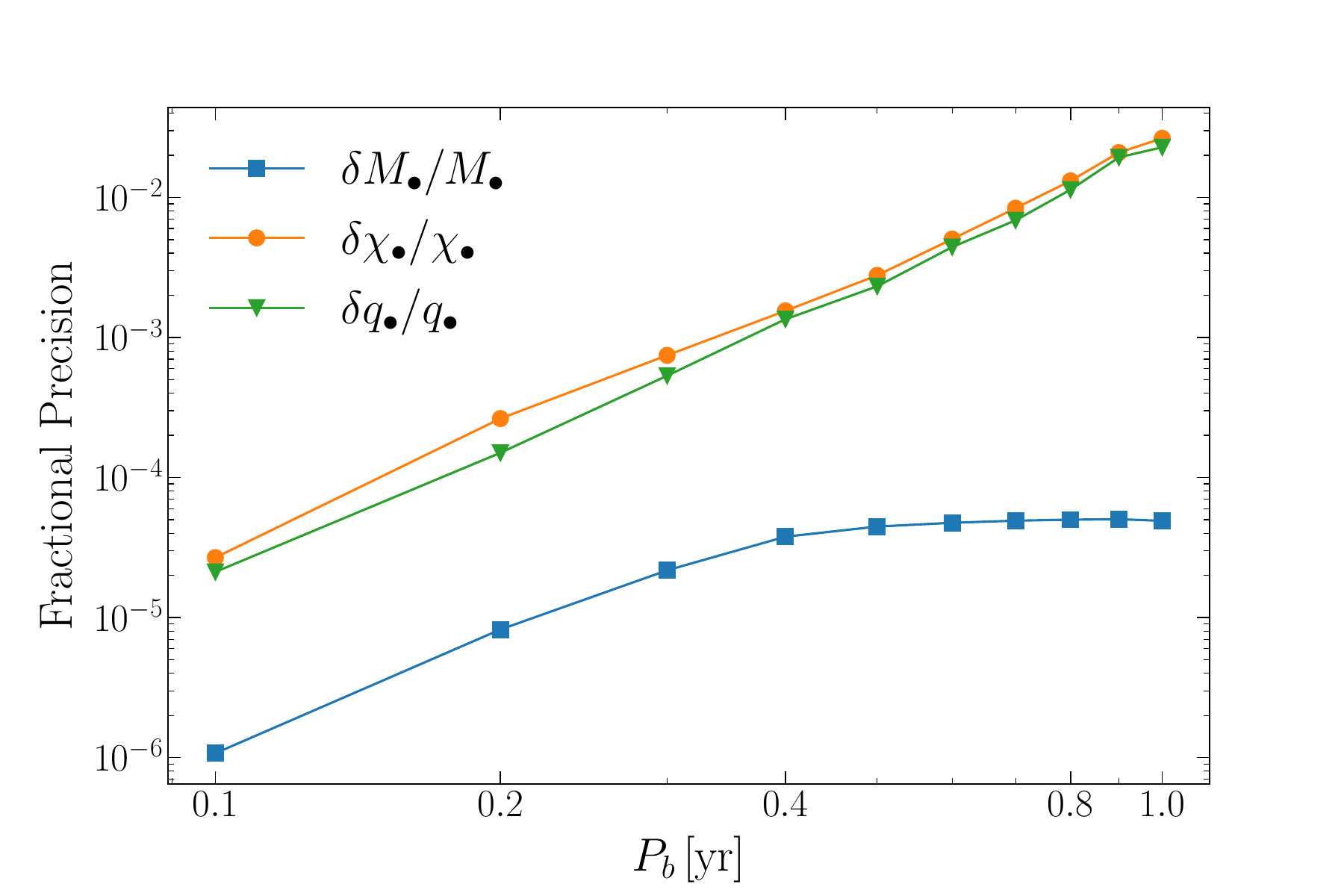}
	\caption{\label{fig:fisher_Pb} Fractional precision for the BH mass, spin
	and quadrupole parameters as functions of the pulsar orbital period. The BH
	parameters and orbital parameters (except $P_b$) are shown in
	Eqs.~(\ref{eq:example:orbit:M}--\ref{eq:example:orbit:angles}). We assume
	that the pulsar spin frequency $\nu=2\,{\rm Hz}$ and the timing precision
	$\sigma_{\rm TOA}=10^{-3}\,{\rm s}$. Simulations include weekly observations
	over a 5-yr interval.}
\end{figure}
%----

Figure~\ref{fig:fisher_Pb} shows the simulation results for the fiducial system
configuration except for the pulsar orbital period which varies from 0.1 yr to
1.0 yr. \ZX{}{The range of $P_b$ chosen here was inspired by \citet{Liu:2011ae}. Previous studies suggest that $\sim 10^3$ pulsars can be expected to be orbiting around Sgr~A* with $P_b<100\,{\rm yr}$~\cite{Wharton:2011dv}, while the innermost some of them could be in orbits as tight as $\sim 100-500\,{\rm au}$ from Sgr~A*~\cite{Zhang:2014kva}, which correspond to the orbital periods considered here.} \ZX{}{ From Fig.~\ref{fig:fisher_Pb}, we} can see that for orbits with $P_b\lesssim 0.5\,{\rm yr}$, the
measurement of the spin and quadrupole parameters can be better than 1\%, which
is consistent with the results in Refs.~\cite{Liu:2011ae, Bower:2018mta}.

%----
\begin{figure}
	\centering
	\includegraphics[width=9.5cm, trim=0 0 0 20]{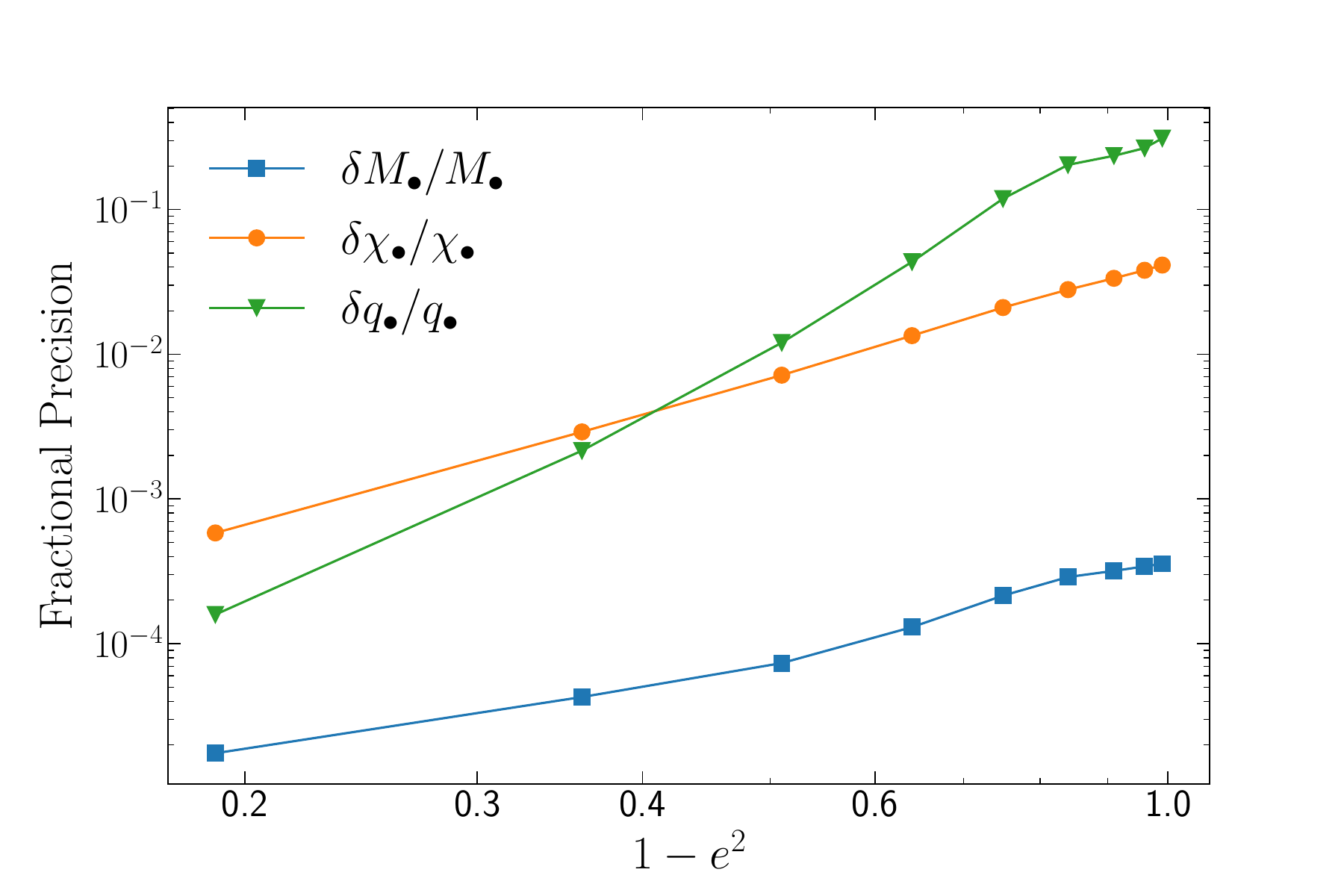}
	\caption{\label{fig:fisher_e}Fractional precision for the BH mass, spin and
	quadrupole parameters as functions of the orbital eccentricity. The system
	parameters (except $e$) are the same as before while the orbital
	eccentricity changes from 0.1 to 0.9.}
\end{figure}
%----

We also investigate the influence of the orbital eccentricity and the results
are presented in Fig.~\ref{fig:fisher_e}. It shows that the fractional
precisions of the mass, spin, quadrupole parameters are nearly linear functions
of the factor $\big(1-e^2 \big)$ in the logarithmic scale. This can be explained
intuitively by that the secular effects of mass, spin, and quadrupole  are
proportional to $\big(1-e^2\big)^{-1}$, $\big(1-e^2\big)^{-3/2}$, and
$\big(1-e^2\big)^{-2}$ respectively~\cite{Wex:1998wt, Barker:1975ae}. From this
figure we can also conclude that for obits with very high eccentricities, say
$e\gtrsim 0.8$, the precision of the quadrupole determination can reach the
precision of the spin determination or even better, which comes from the
$\propto r^{-4}$ nature of the quadrupole interaction.

%----
\begin{figure}
	\centering
	\includegraphics[width=9cm, trim=0 50 0 40]{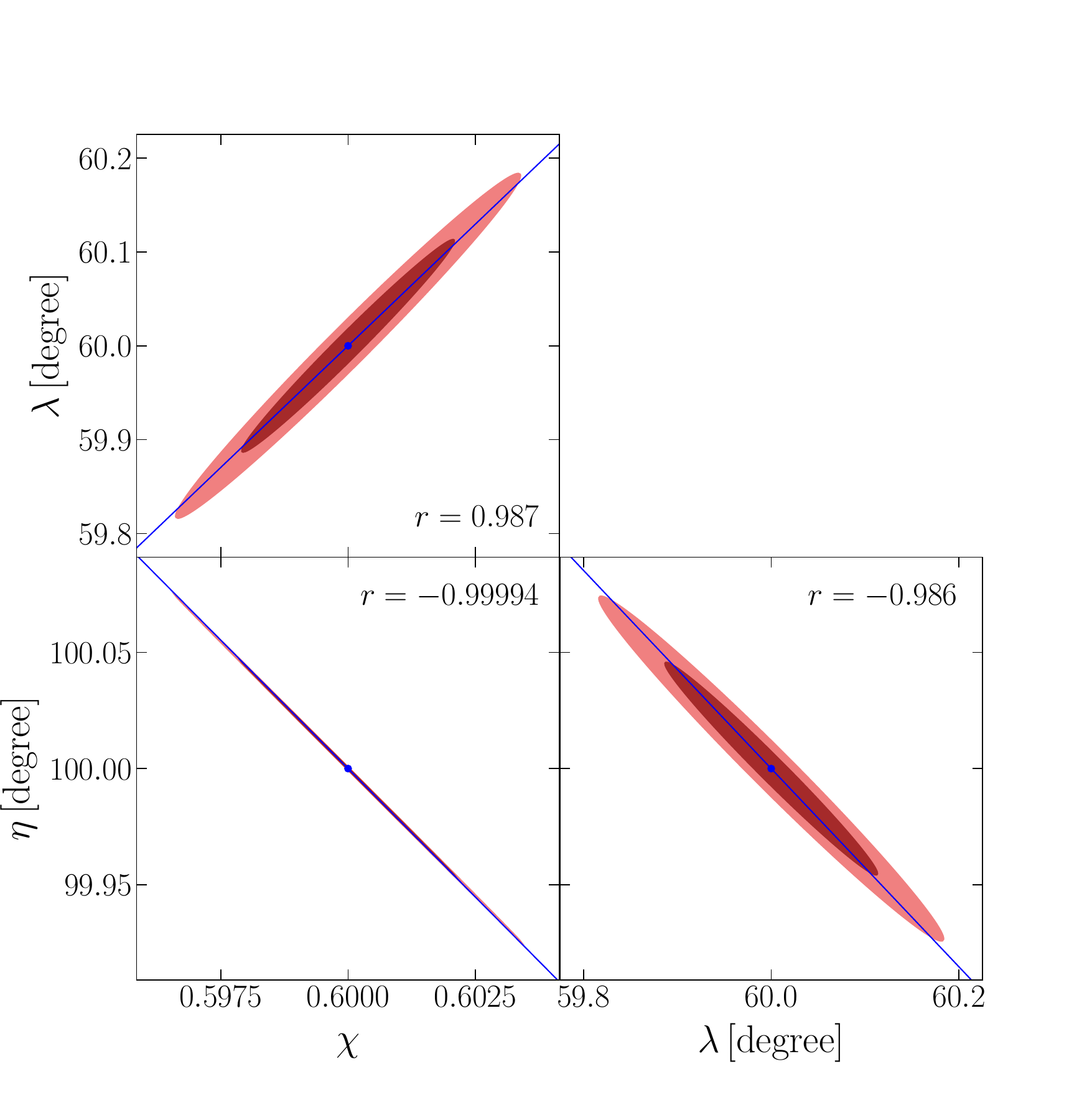}
	\caption{\label{fig:MSQ_cov} Correlations between spin parameters for the
	fiducial system given in
	Eqs.~(\ref{eq:example:orbit:M}--\ref{eq:example:orbit:angles}). The contours
	represent the 68\% and 95\% confidence regions. True values are marked with
	blue dots and the blue lines show the theoretical leading-order degeneracies.}
\end{figure}
%----

As shown in Refs.~\cite{Liu:2011ae, Zhang:2017qbb}, there are leading-order
degeneracies among spin parameters. The observable secular effects caused by the
BH spin including the advance of the periastron $\omega$ and the change of the
inclination angle $i$~\cite{Wex:1998wt}. However, even if one can separate the
periastron advance caused by the Schwarzschild precession, these two secular
effects still  cannot fully determine the three spin parameters. Higher-order
secular effects, for example, the change of the periastron advance rate
$\ddot{\omega}$, or periodic effects are needed when measuring the BH spin and
its orientation~\cite{Liu:2011ae}. In Fig.~\ref{fig:MSQ_cov} we plot the
correlation between the spin parameters, where blue lines give the predicted
leading order degeneracies determined by~\cite{Zhang:2017qbb}
%--
\begin{eqnarray}
	\bold{S}\cdot \big(\hat{\bm{L}}\times\hat{\bm{K}}_0 \big) &=&  {\rm const.}
	\,,\\
	2 \big(\hat{\bm{L}}\cdot\bold{S} \big)+ \big(\hat{\bm{A}}\cdot\bold{S}
	\big)\frac{\cos i}{\sin\omega\sin i} &=& {\rm const.} \,,
\end{eqnarray}
%--
where $\bold{S}=S\hat{\bm{s}}$ is the spin vector of the BH and $\hat{\bm{A}}$
is the unit vector pointing to the periastron from Sgr A*. Here we have used the
initial orbital elements in
Eqs.~(\ref{eq:example:orbit:M}--\ref{eq:example:orbit:angles}) to calculate the
leading-order degeneracies. The small discrepancies between the predicted
directions and the major-axis of the contours may come from the change of the
orbital elements and the higher-order contributions in the precessions or from
the periodic
effects.

%---------------------------------------------------------------------
\section{Dark Matter Perturbation}\label{sec:DM pert}
%---------------------------------------------------------------------

In previous sections we have assumed the system to be sufficiently clean and
estimated the potential of testing gravity using such PSR-SMBH systems. In
reality, there may be various effects that could complicate or even spoil these
tests. Such as the gravity perturbation from massive objects and surrounding
mass distribution~\cite{Merritt:2009ex}. One way to avoid the external
perturbation is to focus on the timing features near the periastron, where the
interaction with central BH is dominating~\cite{Psaltis:2015uza}. In this work
we choose to extend our timing model to include the parameters that describe the
perturbations and study the actual influences of specific kinds of
perturbations. Here we consider the perturbations from a spherically distributed
DM mini-halo around the GC. \ZX{}{The DM profiles are introduced in Sec.~\ref{subsec:NFW} and Sec.~\ref{subsec:Einasto}. In Sec.~\ref{subsec:esti} we present the extended timing model and the parameter estimation results.}

\ZX{}{\subsection{The Generalized NFW Profile} \label{subsec:NFW} }

The standard $\rm \Lambda CDM$ model of the universe has been remarkably
successful in explaining the evolution of the universe and the development of
large-scale structures.  The cold-DM-only simulation gives a nearly
mass-independent DM halo density distribution, the so-called 
NFW profile~\cite{Navarro:1996gj}
%--
\begin{equation}
	\rho_{\rm NFW}(r)=\frac{\rho_0}{(r/R_s)(1+r/R_s)^2}\,,
\end{equation}
%--
where $R_{s}$ is the scale radius and $\rho_0$ is a characteristic density. This
profile has been widely used in describing galaxies' DM halos and fitting for
the mass distribution in lensing observations. For our Milky Way, it gives
$R_{s}\simeq 20\,{\rm kpc}$, and $\rho_0\big|_{r\approx 8\,{\rm kpc}} =0.4\,{\rm
GeV\,cm^{-3}}$ at the location of our Solar System~\cite{McMillan:2017}.

A feature of the NFW profile is the singular density profile at the center where
$\rho(r)\propto r^{-1}$, which is called a density cusp. However, observations
of the rotation curve of low-surface-brightness galaxies suggest that some
galaxies have a finite density core which can form in some other DM
models\ZX{}{ such as the self-interacting DM (SIDM) model}~\cite{Moore:1999gc,Vargya:2021qza}. Although the cold DM model has been successfully
examined by observations at large scales, the small scale properties of DM are
still lack of constraints~\cite{Hui:2016ltb}. Thus a measurement of the central
profile of DM density may help us understand it better and constrain different
DM models.  Various baryonic processes such as adiabatic contraction or density
fluctuations due to supernova feedback could have modified the DM
profile~\cite{Pontzen:2011ty,Cole:2011yp}\ZX{}{, while recent simulations also suggest that SIDM models do not produce large differences in the inner structure of Milky-Way-mass galaxies in the presence of baryonic feedback effects~\cite{Vargya:2021qza}.}  \ZX{}{ Nevertheless, to} account for different model
predictions, here we consider a generalized model, the gNFW
model~\cite{Diemand:2008in}
 %--
\begin{equation}
	\rho_{\rm gNFW}(r)=\frac{\rho_0}{(r/R_s)^\gamma(1+r/R_s)^{3-\gamma}}\,.
\end{equation}
%--
The central behaviors of different models are characterized by the power-law
index $\gamma$. 

\citet{Gondolo:1999ef} pointed out that the adiabatic growth of the SMBH in the
DM halo center will strongly modify the DM profile inside the radius of its
gravitational influence, $R_h \equiv GM_\bullet/v_0^2$, which is about
$1.7\,{\rm pc}$ for Sgr A*~\cite{Gultekin:2009qn}. For collisionless DM with a
polytrope phase-space distribution, a DM spike with $\rho_{\rm sp}(r)\propto
r^{-\gamma_{\rm sp}}$ will form, with $\gamma_{\rm sp}=(9-2\gamma)/(4-\gamma)$.
For a pulsar with orbital period $P_b \sim 1\,{\rm yr}$ has a semi-major axis
$\sim 1\,{\rm mpc}$, which is at the region that is dominated by the central BH.
It motivates us to take this effect into account.

Some DM models allow the DM particles to annihilate with
themselves~\cite{Gondolo:1999ef}. If so, due to the very high DM densities in
the halo center, it may act as a strong gamma-ray source~\cite{Gondolo:1999ef},
and the spike induced by the SMBH will enhance it further. On the other hand,
the annihilation of DM particles will limit the maximum density of the
DM spike, and produce a weak cusp in the center with $\rho_{\rm in}(r)\propto
r^{-\gamma_{\rm in}}$, where $\gamma_{\rm in}\simeq 0.5$~\cite{Vasiliev:2007vh}. Such
a weak annihilation cusp will form around a radius $R_{\rm in}$ where the DM
density reaches a critical density $\rho_{\rm ann}$.

Combining the above discussion, we consider the DM density profile as
below~\cite{Shao:2018klg},
%--
\begin{equation}
	\rho_{\rm DM}(r)=\left\{
	\begin{aligned}
		&\frac{\rho_{\rm sp}(r)\rho_{\rm in}(r)}{\rho_{\rm sp}(r)+\rho_{\rm
		in}(r)}\,,&4GM_{\bullet}/c^2 \leq r<R_{\rm sp}\\
		&\rho_{\rm gNFW}(r)\,,&r\geq R_{\rm sp}\ \ \ \ \ \ \ \ \ \ \ \ \ \ \ \ \ \ \ 
	\end{aligned}
	\right.\,.
\end{equation}
%--
If there is no DM annihilation, the profile is 
%--
\begin{equation}\label{eq:spike den}
	\rho_{\rm DM}(r)=\left\{
	\begin{aligned}
		&\rho_{\rm sp}(r)&4GM_{\bullet}/c^2 \leq r<R_{\rm sp}\\
		&\rho_{\rm gNFW}(r)\,,&r\geq R_{\rm sp}\ \ \ \ \ \ \ \ \ \ \ \ \ \ \ \ \ \ \ 
	\end{aligned}
	\right.\,,
\end{equation}
%--
where $R_{\rm sp}$ is chosen to equal to the gravitational radius $R_h$.
Densities $\rho_{\rm sp}(r)$ and $\rho_{\rm in}(r)$ are determined by the
continuous condition at $R_{\rm sp}$. For the model with annihilation, we assume
$\rho_{\rm sp}(R_{\rm in})=\rho_{\rm in}(R_{\rm in})=2\rho_{\rm ann}$ to
interpolate these two profiles. 

%--------------------------
\begin{figure}
	\centering
	\includegraphics[width=8.5cm, trim=10 120 10 100]{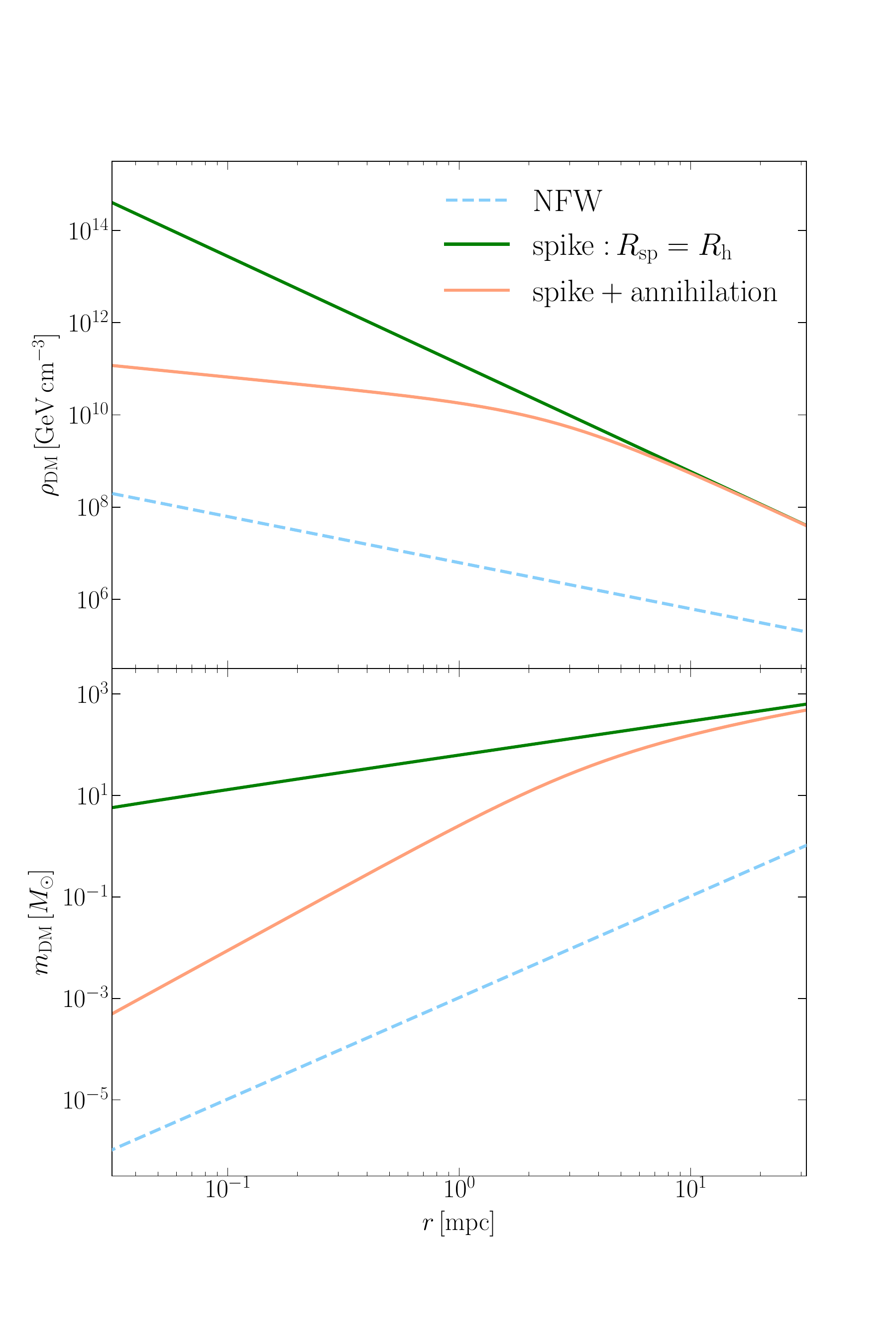}
	\caption{\label{fig:DM distribution}The DM density profile (upper) and the
	DM mass inside $r$ (lower) for the NFW model, the NFW model with DM spike
	and the model with DM annihilation.}
\end{figure}
%--------------------------

\ZX{}{\subsection{The Einasto Profile} \label{subsec:Einasto}}

\ZX{}{The Einasto profile is also a commonly used DM distribution in which the logarithmic density slope shows a power-law behavior~\cite{Einasto:1965czb,Navarro:2003ew,Merritt:2005xc}, which cannot be characterized by the gNFW model we discussed before. It is argued that this model provides a better fit to the high-resolution $N$-body DM simulation and it gives a core-like central behavior~\cite{Navarro:2003ew,Wang:2019ftp}. The DM density profile of this model can be written as~\cite{shen2023exploring}
\begin{equation}
    \rho_{\rm E}(r)=\rho_0\exp\left\{-\frac{2}{\alpha}\left[\left(\frac{r}{r_s}\right)^\alpha-1\right]\right\}\,,
\end{equation}
where $\rho_0$ is the DM density at the scale radius $r_s$ and $\alpha$ is the inverse of the Einasto index. Fitting the rotation curve and globular cluster kinematics from \textit{Gaia} data gives~\cite{Wang:2021nkz}
\begin{align}
  \rho_{\rm E}(8.2\,{\rm kpc}) &=0.008\,M_\odot\,{\rm pc^{-3}} \,, \\
  r_s &=12\,{\rm kpc} \,, \\
  \alpha &=0.32 \,.
\end{align}

The analytic form of the DM spike induced by the adiabatic growth of the SMBH in the Einasto model is not available in the literature. However, one may estimate it with circular-orbit approximation~\cite{shen2023exploring}. Following \citet{shen2023exploring}, we estimate the DM spike via
\begin{align}
    r_iM_{{\rm tot},\,i}(r_i)&=r_fM_{{\rm tot},\,f}(r_f)\,,\\
    M_{{\rm DM},\,i}(r_i)&=M_{{\rm DM},\,f}(r_f)\,,
\end{align}
where $M_{\rm DM}(r)$ is the mass of DM enclosed in the radius $r$
\begin{equation}
    M_{\rm DM}(r)=\int_0^r 4\pi r^2 \rho_{\rm DM}(r){\rm d}r\,,
\end{equation}
and $M_{\rm tot}(r)=M_{\rm SMBH}+M_{\rm DM}(r)$. Exact distribution of the DM
spike in the Einasto model can be numerically calculated as shown in
Ref.~\cite{shen2023exploring}. Considering that we are only interested in the
central behavior of the DM spike at the radius scale of mpc, which is much
smaller than the scale radius $r_s$, it is a good approximation to have
$\rho_E(r)=\rho_0e^{2/\alpha}$. Combining with above equations, one derives that
the DM spike density in the Einasto model is $\rho_{\rm E,\,sp}(r)\propto
r^{9/4}$, which is effectively to have $\gamma=0$ in the gNFW model. It is also
demonstrated that even the initial Einasto profile is much smaller than the NFW
profile at the central region, the DM spike densities in these two model in fact
can have very similar orders of magnitude~\cite{shen2023exploring}. Even though
the Einasto profile itself cannot be desicribed by the gNFW profile, we conclude
that the DM spike distribution in the Einasto model can be well characterized by
the first expression of Eq.~(\ref{eq:spike den}), and two models will give
basically the same results.

}

In Fig.~\ref{fig:DM distribution} we show the DM density profile and the DM mass
enclosed by the radius $r$. We take the NFW model as an example and show the
related models with spike and DM annihilation. Using the method of osculating
elements~\cite{poisson_will_2014}, one can calculate the secular effects caused
by the DM distribution. Due to the spherical symmetry, the only secular effect
caused by the DM is the modified advance of periastron.
%--------------------------
\begin{figure}
	\centering
	\includegraphics[width=9cm, trim=0 20 15 20]{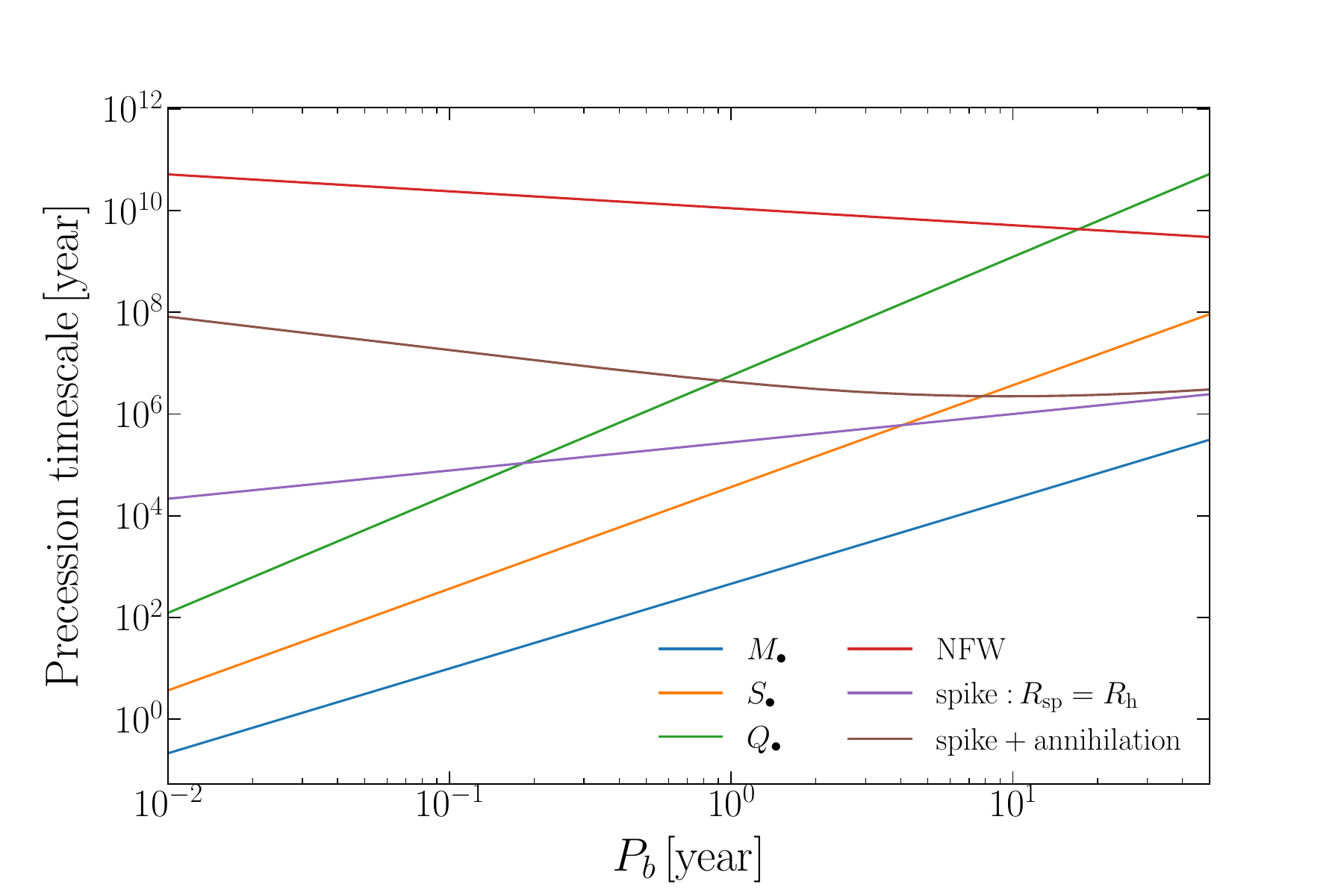}
	\caption{\label{fig:precession}Precession time scales as functions of the
	orbital period for various relativistic effects and the DM perturbation.}
\end{figure}
%--------------------------
The precession time scales of various effects in PSR-Sgr~A* systems are shown in
Fig.~\ref{fig:precession}. From Fig.~\ref{fig:DM distribution} and
Fig.~\ref{fig:precession}, we can see that the DM mass contributed by the
initial NFW model is far below the precision of BH mass determination obtained
before, which is about $10$--$100\,M_\odot$, while for the DM model with
annihilation, timing a pulsar with an orbital period $P_b\lesssim 10\,{\rm yr}$
can only give the information of the weak cusp caused by the DM annihilation,
described by $\gamma_{\rm in}$, which is not strongly related to the original DM
profile. For a pulsar with a larger orbital period, in principle one may have a
measurement of $\gamma_{\rm sp}$, but the complex environments and external
perturbations will complicate this system. For DM model without annihilation,
the spike structure induced by the BH can largely increase the DM density, thus
timing a pulsar with a reasonable orbital period will give a constraint on
$\gamma_{\rm sp}$, which is related to the underlying DM profile via
$\gamma_{\rm sp}=(9-2\gamma)/(4-\gamma)$.

\ZX{}{\subsection{Prameter Estimation} \label{subsec:esti}}

\begin{figure}
	\centering
	\includegraphics[width=9cm, trim=0 10 15 10]{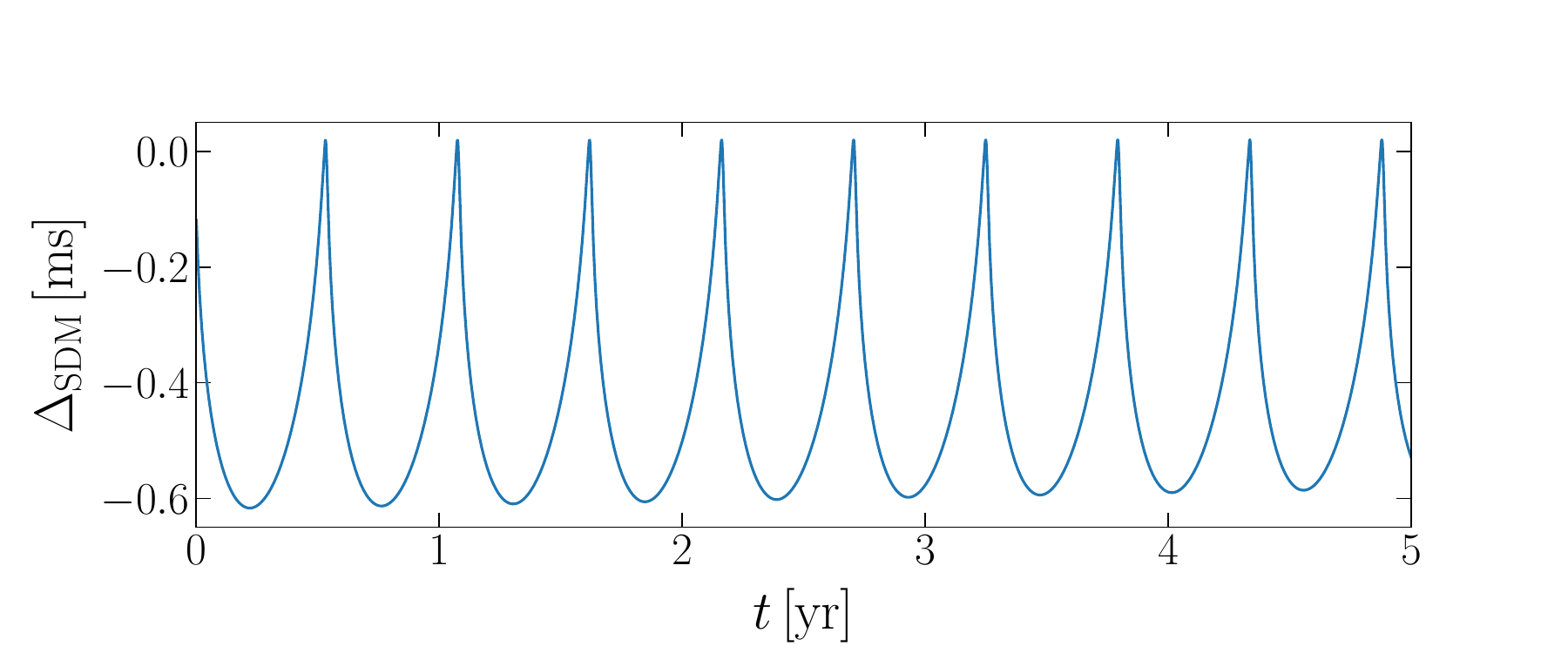}
	\caption{\label{fig:Delta_SDM} The Shapiro time delay caused by the DM
	distribution in the spike DM model which is based on the NFW profile. The
	system parameters are given in
	Eqs.~(\ref{eq:example:orbit:M}--\ref{eq:example:orbit:angles}).}
\end{figure}

Compared to the mass of the central BH, the DM mass is still very small (about
$10^{-5}\,M_\bullet$ for the spike model inside 1\,mpc). Thus we may treat the
DM as a perturbation, which means that we only consider the Newtonian gravity
caused by the DM distribution and find the effects on the pulsar's orbit. To
verify this assumption, as an example, we calculate the leading-order Shapiro
time delay caused by the extended DM mass distribution in the spike model, as
shown in Fig.~\ref{fig:Delta_SDM}, which is for a system with parameters in
Eqs.~(\ref{eq:example:orbit:M}--\ref{eq:example:orbit:angles}).  One can simply
estimate from Fig.~\ref{fig:time:delays} that the Shapiro time delay caused by
the DM should be at the order of 1 ms, which is consistent with the calculation,
while the extended mass distribution slightly weakens the amplitude and broadens
its shape. The Shapiro delay caused by the DM is still almost degenerate with
the Shapiro delay of the BH and its value is smaller than the assumed timing
precision. Even with a larger orbit that includes more DM mass, we can still
ignore this contribution in our setting.

%--------------------------
%-------------------------

%--------------------------
\begin{figure}
	\centering
	\includegraphics[width=9cm, trim=0 0 10 20]{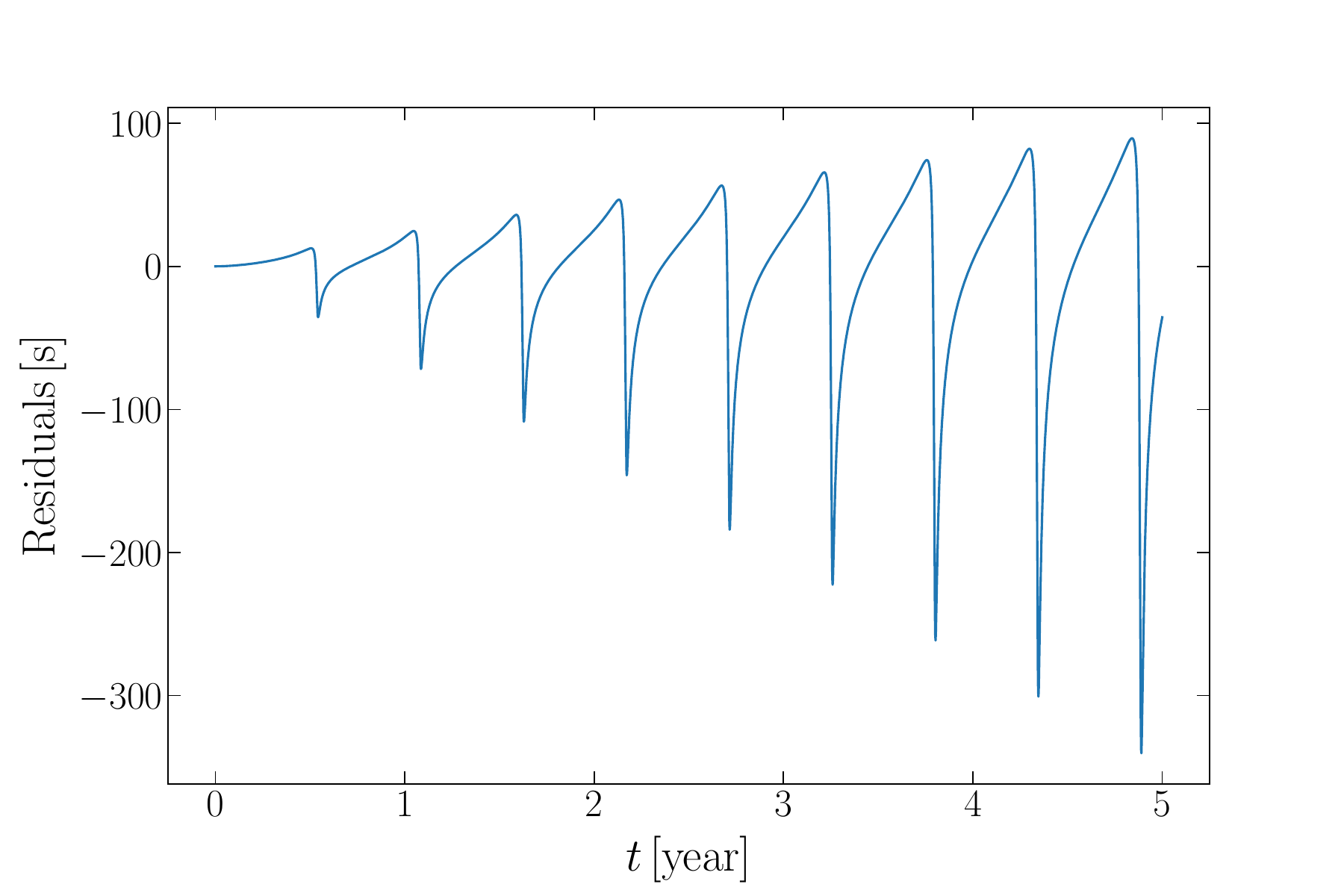}
	\caption{\label{fig:spike_before}Timing residuals caused by the DM
	perturbation before fitting. We use the DM model with spike but no
	annihilation, and we set $R_{\rm sp}=R_h$. The system parameters are given
	in Eqs.~(\ref{eq:example:orbit:M}--\ref{eq:example:orbit:angles}).}
\end{figure}
%--------------------------

%--------------------------
\begin{figure}
	\centering
	\includegraphics[width=9cm, trim=0 0 10 10]{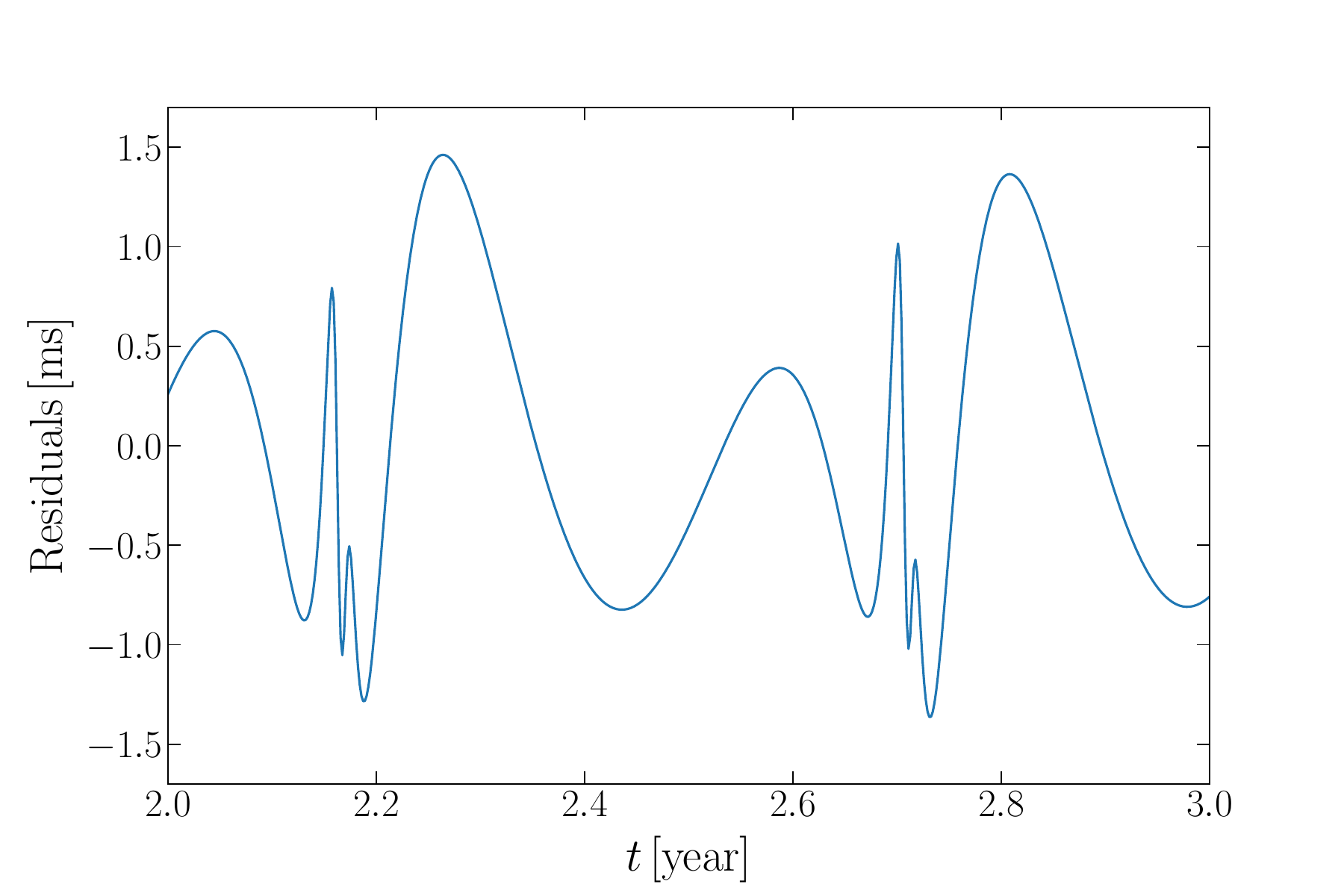}
	\caption{\label{fig:spike_after}Same as Fig.~\ref{fig:spike_before}, but
	after fitting for a model without DM. Most of the residuals in
	Fig.~\ref{fig:spike_before} are absorbed in other parameters.}
\end{figure}
%----------------------------

In Fig.~\ref{fig:spike_before} and Fig.~\ref{fig:spike_after} we present the
timing residuals caused by the DM perturbation. We simulated a set of TOAs with
the full timing model that includes the DM effects and fitted for the TOAs with
the model described in Sec.~\ref{sec:pulsar timing}\ZX{}{, which does not include the DM distribution}. \ZX{}{This is the case for real observations that in the beginning people usually use a simple timing model that accounts for those largest effects as a first step.} Figure~\ref{fig:spike_before} shows the residuals before fitting, i.e. we use
the true parameters $\bar{\bm{\Theta}}$ that are used in simulation but do not
include the DM contributions. \ZX{}{Although the true parameters in principle are unknow, this figure shows the cumulated timing residuals caused by the DM perturbation.} There is a secular part in the residuals due to
the periastron advance caused by the DM perturbation.
Figure~\ref{fig:spike_after} shows the residuals after fitting. The secular part
of the residuals is mainly absorbed by the BH spin parameter and only a
quasi-periodic part is left. \ZX{}{ We plot the residuals for $t$ from the second year to the third year of the total observation time span, which corresponds to two orbital periods of the pulsar.} The
residuals after fitting remain an amplitude of about 1\,ms, which is close to
the assumed timing precision, showing a possibility of estimating the DM
parameters with pulsar timing. \ZX{}{The timing residuals for the initial NFW model and the spike model with DM annihilation have similar behaviors as show in Fig.~\ref{fig:spike_before} and Fig.~\ref{fig:spike_after}, but with a much smaller amplitude as discussed above.} The observation of S2 orbit has set a limit on
the extended mass inside the S2's apocenter, which is about
$3000\,M_\odot$~\cite{GRAVITY:2021xju, Heissel:2021pcw}, and the DM model we
considered here gives a value that is consistent with this constraint. The
extended mass inside the S2 orbit contributed by the DM in the spike model is
$\lesssim 1000\,M_\odot$, although the star clusters will also contribute to the
extended mass in this scale~\cite{Heissel:2021pcw}. 

As discussed before, we extend our timing model by adding a DM distribution
related to the DM spike structure, which is a density profile with two
parameters
%--
\begin{equation}\label{eq:DM profile}
	\rho_{\rm DM}(r)=\rho_0\left(\frac{r}{4GM_\bullet/c^2}\right)^{-\gamma_{\rm
	sp}}\,,\ \ r\geq 4GM_\bullet/c^2\,,
\end{equation}
%--
where $\rho_0$ now represents the central density of the DM spike and
$\gamma_{\rm sp}$ is related to the original DM profile, where we use the NFW
profile with $\gamma=1$ in our simulation. 

%--------------------------
\begin{figure}
	\centering
	\includegraphics[width=9.2cm, trim=0 0 10 10]{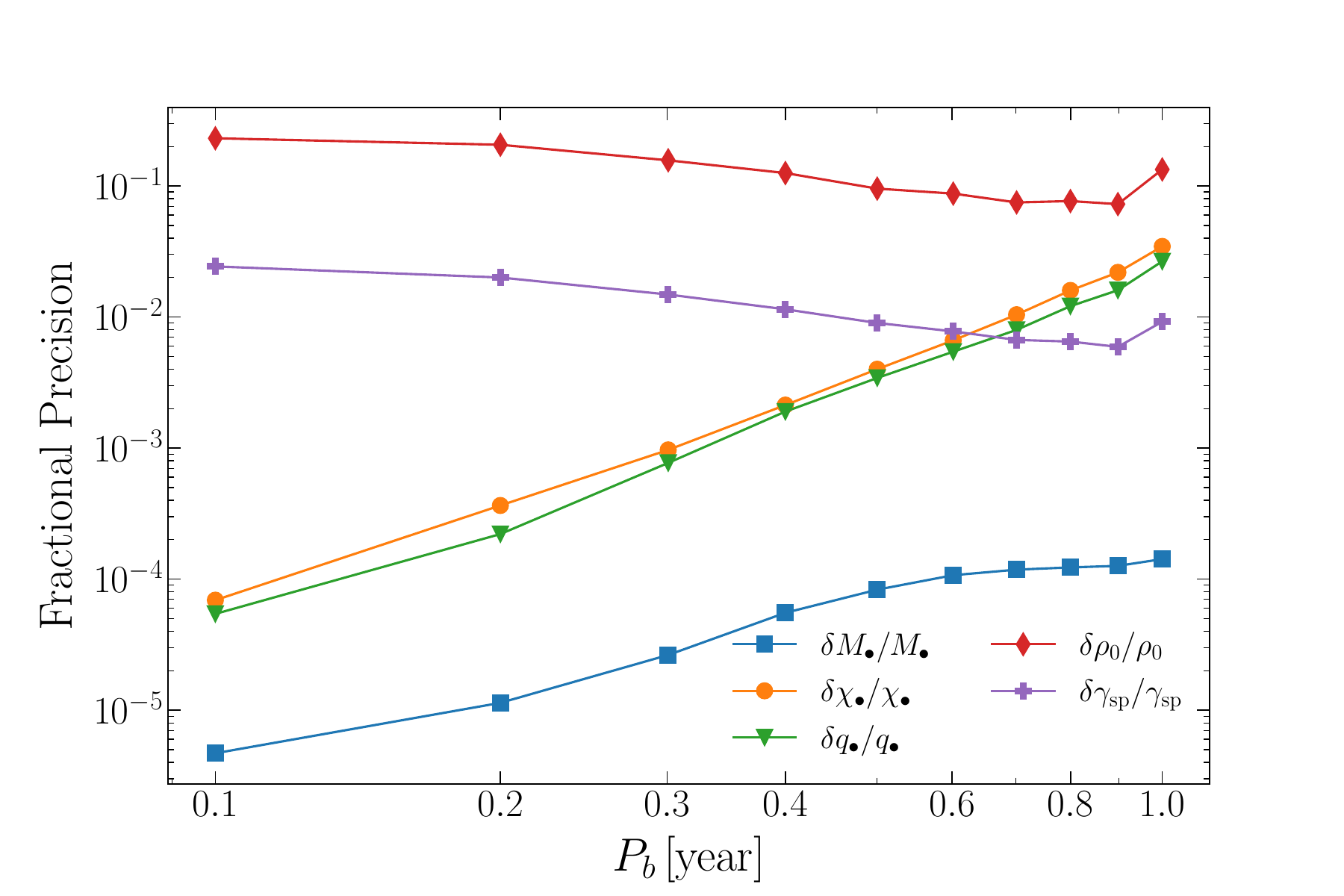}
	\caption{\label{fig:DM_estimate_Pb} Fractional precision for the BH mass,
	spin, and quadrupole parameters, as well as the DM parameters. Parameters
	for Sgr A* and the pulsar orbit are shown in
	Eqs.~(\ref{eq:example:orbit:M}--\ref{eq:example:orbit:angles}). The
	parameter $\rho_0$ and $\gamma_{\rm sp}$ are the parameters for the
	NFW model with spike.}
\end{figure}
%--------------------------

We fit the TOAs with the full model where the effects from DM are included.  In
Fig.~\ref{fig:DM_estimate_Pb} we show the results of parameter estimation. The
input parameters of the BH and the pulsar are shown in
Eqs.~(\ref{eq:example:orbit:M}--\ref{eq:example:orbit:angles}), and the DM model
is the NFW model with spike. We still assume weekly observations over a 5-yr
interval. We can see that timing a pulsar with an orbital period $P_b\gtrsim
0.5\,{\rm yr}$ and an orbital eccentricity $e\sim 0.8$ can give a fractional
measurement uncertainty of $\sim 1\%$ in $\gamma_{\rm sp}$ for the spike model,
which is related to a $\sim 20\%$ fractional precision in $\gamma$. This result
is comparable to the constraints from fitting kinematic data of maser and other
observations, which give $\gamma=0.79\pm0.32$ for our
Galaxy~\cite{McMillan:2017}.  However, by timing a pulsar around Sgr~A*, we can
constrain the DM structure in the length scale of $\sim {\rm mpc}$, which is 6
orders of magnitude smaller than the length scale of other Galactic observations
of such an investigation, which are typically done at $\sim {\rm kpc}$ scales.

%--------------------------
\begin{figure}
	\centering
	\includegraphics[width=9.2cm, trim=0 15 10 10]{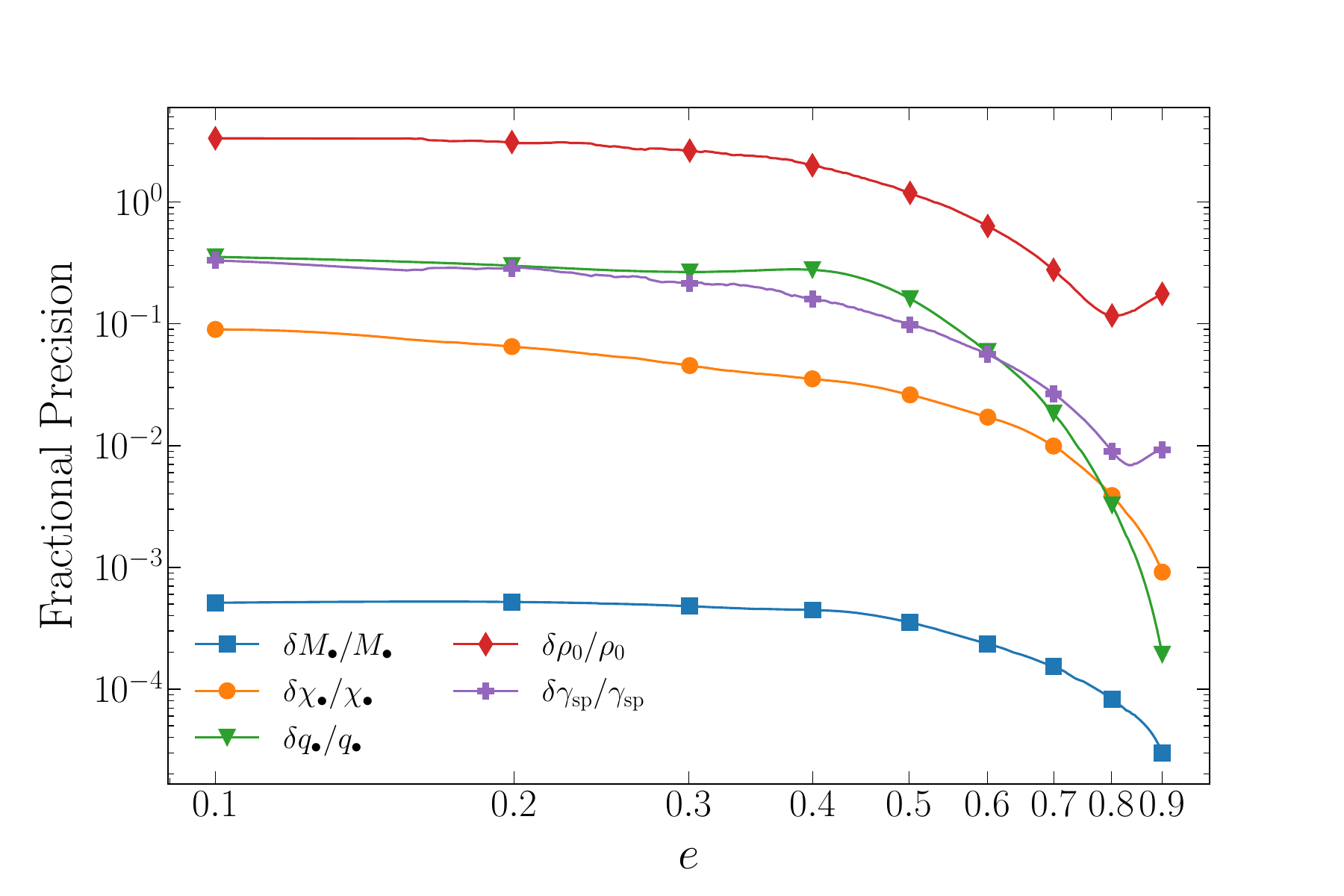}
	\caption{\label{fig:DM_estimate_e} Similar to Fig.~\ref{fig:DM_estimate_Pb},
	but as a function of the orbital eccentricity.}
\end{figure}
%--------------------------

We also investigate the effects of the orbital eccentricity and the result is
shown in Fig.~\ref{fig:DM_estimate_e}. We can see that the measurement
precisions of the DM parameters decrease fast as the orbital eccentricity
becomes small. This is because for small orbital eccentricity, the effect caused
by the DM distribution is strongly degenerate with the BH mass, providing only a
mass monopole in the limit of $e\to 0$. Only when the pulsar moves at different
radii it can sense the DM's radial distribution.

In the above, we have used unrelated $\chi_\bullet$ and $q_\bullet$, although in
GR they can be related via Eq.~(\ref{eq:nohairtheorem}).  Another consideration
is to detect the DM profile under the assumption that GR is correct, which
corresponding to setting $q_\bullet=-\chi_\bullet^2$ in the timing model. The
constraints on the DM parameters obtained under this assumption are similar to
what are shown in Fig.~\ref{fig:DM_estimate_Pb} and
Fig.~\ref{fig:DM_estimate_e}.  This can be partly explained by
Fig.~\ref{fig:precession}. As the only secular effect caused by the DM
distribution is the periastron advance, the leading-order degeneracy among the
DM parameters and other parameters are dominated by this secular effect, which
is similar to the discussions on the leading-order degeneracy among the spin
parameters. Thus the quadrupole effect can only have a very small contribution
compared to the others as shown in Fig.~\ref{fig:precession}.  Also it has been
discussed by \citet{Heissel:2021pcw} that, the orbital features caused by the
extended mass distribution are significant in the orbital section
$\theta\in(5\pi/6,\,7\pi/6)$, which is different from the spin or quadrupole
effects that are most significant near the periastron. So setting
$q_\bullet=-\chi_\bullet^2$ only has small effect on the DM measurement.
However, if the central object is totally different from GR, such as if it is a
supermassive boson star that can have larger quadrupole moment, the conclusion
may be different.

%---------------------------------------------------------------------
\section{Conclusions}\label{sec:conclusion}
%---------------------------------------------------------------------

In this work, we explore the prospects of constraining the BH properties and DM
models from timing a pulsar around Sgr~A*. We construct a timing model based on
the numerical integration of the PN equation of motion, with leading-order
effects of various time delays being taken into account. Our simulations show
that for a pulsar with an orbital period $ P_b \lesssim 0.5\,{\rm yr}$ and an
orbital eccentricity $e \sim 0.8$, a 5-yr observation with weekly recorded TOAs
and a timing precision 1\,ms, the BH spin and quadrupole parameters can be
measured with a precision of $10^{-2}$ or better, which is consistent with
previous studies based on a semi-analytic approach~\cite{Liu:2011ae,
Psaltis:2015uza} or a fully general-relativistic treatment~\cite{Zhang:2017qbb}.
We need to emphasize that the timing model in this work only considers the
leading-order effects, which is sufficient for estimating the measurement
precision of the system parameters but not enough for applying to the real
observations. Higher-order effects should be carefully considered and added into
the timing model if the observational precision is high enough. We hope to
further develop the timing model along this line in future studies.

As a concrete application---which is hard to achieve with the semi-analytic
approach or the fully  general-relativistic treatment---by extending our timing
model with a spherical DM perturbation, we investigate the measurement precision
of the DM distribution at small scales. For DM models with a spike structure
induced by the adiabatic growth of the central BH, timing a pulsar with an
orbital period $P_b \gtrsim 0.5\,{\rm yr}$, and an orbital eccentricity $e \sim
0.8$ can provide a 1\% measurement of the power-law index $\gamma_{\rm sp}$ of
the DM spike, which relates to a measurement of the underlying DM model with a
$~20\%$ precision in the index $\gamma$.  Such a precision is comparable with,
but complementary to, the observations at the Galactic scale, usually done at
$\sim$ kpc scales\ZX{}{, which can be an equilibrium based kinematic analysis~\cite{McMillan:2017}, or some extreme-precision time-series measurements of Galactic accelerations~\cite{Chakrabarti:2020kco,Chakrabarti:2021ypi}}. As well known, measuring the DM
distribution is in general easier in larger scales, where the DM can contribute
a larger total mass. However, the various environmental effects can complicate
the test in this situation. With future discovery of proper pulsars near the GC,
pulsar timing observation with high precision will provide us a unique
opportunity to explore the small-scale properties of the DM, and eventually lead
to a more complete understanding of the origin of the DM.

%---------------------------------------------------------------------

%---------------------------------------------------------------------
\acknowledgments
%---------------------------------------------------------------------

We thank Norbert Wex for helpful discussions, and Kuo Liu for carefully reading
the manuscript.  This work was supported by the National SKA Program of China
(2020SKA0120300), the National Natural Science Foundation of China (11991053,
11975027, 12273006), the Max Planck Partner Group Program funded by the Max
Planck Society, and the High-Performance Computing Platform of Peking
University. 

% 
%---------------------------------------------------------------------
\bibliography{refs.bib}
%---------------------------------------------------------------------

\end{document}